\documentclass[12pt]{article}

\usepackage{lipsum}
\usepackage{multirow}
\usepackage[utf8]{inputenc}
\usepackage{xcolor, graphicx}
\usepackage{framed}
\usepackage{enumitem}
\newlist{steps}{enumerate}{1}
\setlist[steps, 1]{label = \emph{Step \arabic*}:}
\usepackage{makecell}
\usepackage{tikz}
\usepackage{mathpazo}
\usepackage{arydshln} % for dashed line https://tex.stackexchange.com/questions/20140/can-a-table-include-a-horizontal-dashed-line
\usepackage[flushleft]{threeparttable} % http://ctan.org/pkg/threeparttablehttps://www.overleaf.com/project/62e25dc82f019c08e13ce2f2
\usepackage{booktabs,caption}
\usepackage{graphicx}
\usepackage{tabularx} % in the preamble
\usepackage[ruled]{algorithm2e}

\definecolor{darkgreen}{rgb}{0.0, 0.26, 0.15}

\usepackage[colorlinks,citecolor=blue,urlcolor=blue,linkcolor=darkgreen]{hyperref}
\usepackage[citestyle=apa,style=apa,backend=biber,url=false,uniquename=false, useprefix=true, doi=false]{biblatex}
\usepackage{dsfont} % for fancy R
\AtEveryBibitem{%
  \clearfield{note}%
}
\usepackage{mathtools, amsfonts, amssymb}
\usepackage{amsmath}
\numberwithin{equation}{section}
\mathtoolsset{showonlyrefs=true}
\allowdisplaybreaks[2]
\usepackage{siunitx}

\providecommand{\keywords}[1]
{
  \small	
  \textit{Some key words---} #1
}

\usepackage{numprint}
\usepackage{booktabs}

\usepackage{chngcntr}
\usepackage{apptools}
\AtAppendix{\counterwithin{lemma}{section}}

\newcommand{\pkg}[1]{{\normalfont\fontseries{b}\selectfont #1}} \let\proglang=\textsf 

\newcommand{\boldbeta}{\boldsymbol{\beta}}

\newcommand{\Expec}{\mathbb{E}}

\newcommand{\boldpsi}{\boldsymbol{\psi}}
\newcommand{\boldy}{\mathbf{y}}

\newcommand{\Var}{\operatorname{Var}}
\newcommand{\Cov}{\operatorname{Cov}}
\newcommand{\diag}{\operatorname{diag}}

\newcommand*\samethanks[1][\value{footnote}]{\footnotemark[#1]}

\title{A Multivariate Multilevel Longitudinal Functional Model for Repeatedly Observed Human Movement Data}
\author{Edward Gunning
\thanks{\textbf{Corresponding author: \href{mailto:edward.gunning@pennmedicine.upenn.edu}{edward.gunning@pennmedicine.upenn.edu}}}
\thanks{Department of Biostatistics, Epidemiology and Informatics, University of Pennsylvania}
\and
Steven Golovkine\thanks{MACSI, Department of Mathematics and Statistics, University of Limerick, Ireland}
\and
Andrew J. Simpkin\thanks{School of Mathematical and Statistical Sciences, University of Galway, Ireland}
\and
Aoife Burke\samethanks[6]
\and
Sarah Dillon\samethanks[6] \samethanks[7] \thanks{School of Allied Health, Faculty of Education and Health Science, University of Limerick, Limerick, Ireland}
\and
Shane Gore\samethanks[6] \samethanks[7]
\and
Kieran Moran\thanks{Centre for Injury Prevention and Performance, Athletic Therapy and Training; School of Health and Human Performance, Dublin City University, Dublin, Ireland}
\thanks{Insight SFI Research Centre for Data Analytics, Dublin City University, Dublin, Ireland}
\and
Siobhan O'Connor\samethanks[6]
\and
Enda Whyte\samethanks[6]
\and
Norma Bargary\samethanks[3]
}
\date{\today}

\begin{document}

\maketitle

\begin{abstract}
 Biomechanics and human movement research often involves measuring multiple kinematic or kinetic variables regularly throughout a movement, yielding data that present as smooth, multivariate, time-varying curves and are naturally amenable to functional data analysis.
It is now increasingly common to record the same movement repeatedly for each individual, resulting in curves that are serially correlated and can be viewed as longitudinal functional data.
In this work, we present a new approach for modelling multivariate multilevel longitudinal functional data, with application to kinematic data from recreational runners collected during a treadmill run.
For each stride, the runners' hip, knee and ankle angles are modelled jointly as smooth multivariate functions that depend on subject-specific covariates.
Longitudinally varying multivariate functional random effects are used to capture the dependence among adjacent strides and changes in the multivariate functions over the course of the treadmill run.
A basis modelling approach is adopted to fit the model -- we represent each observation using a multivariate functional principal components basis and model the basis coefficients using scalar longitudinal mixed effects models.
The predicted random effects are used to understand and visualise changes in the multivariate functional data over the course of the treadmill run.
In our application, our method quantifies the effects of scalar covariates on the multivariate functional data, revealing a statistically significant effect of running speed at the hip, knee and ankle joints.
Analysis of the predicted random effects reveals that individuals' kinematics are generally stable but certain individuals who exhibit strong changes during the run can also be identified.
A simulation study is presented to demonstrate the efficacy of the proposed methodology under realistic data-generating scenarios.

\end{abstract}


% ---------------------------------------------------------------------------
% MAIN PART
\section{Introduction}

\emph{Longitudinal functional data analysis} (LFDA) concerns the analysis of functional data (e.g., curves or images) that are collected in a longitudinal study design, i.e., functions are collected at repeated time points for multiple subjects \parencite{park_longitudinal_2015}. Examples include daily activity functions measured consecutively for a number of days for several subjects \parencites{goldsmith_generalized_2015}{li_fixed-effects_2022} or brain imaging profiles of patients measured at several hospital visits \parencite{greven_longitudinal_2010} (see \textcite{park_longitudinal_2015}). In contrast to the use of functional data analysis to model sparse and irregular scalar measurements observed longitudinally, an area which has received significant attention \parencite{zhao_functional_2004, rice_functional_2004, muller_functional_2005, yao_functional_2005}, LFDA is concerned with modelling dependence among \emph{functions} due to correlation over a longer (or different) timescale than the one on which they are measured. 

\begin{figure}[h]
    \centering
    \includegraphics[page = 1, width = 1\textwidth]{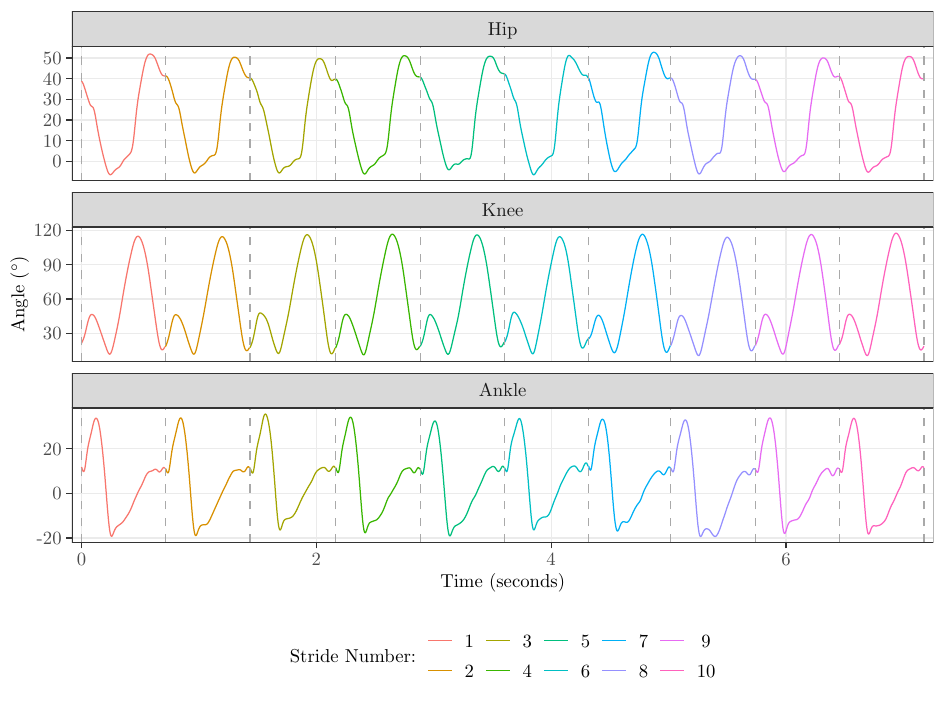}
    \caption{The right sagittal hip, knee and ankle angles of a single participant in the RISC dataset for the first ten strides of their treadmill run. The dashed vertical lines indicate touch down (i.e., when the foot first touches the ground), which represents the start and end of each stride.}
    \label{fig:sample-adjacent-strides-sub-01}
\end{figure}

Our motivating dataset comes from the Dublin City University running injury surveillance (RISC) study, where kinematic data from recreational runners were captured during a treadmill run with the goal of understanding running technique and its link to injury.
In this work, we focus on modelling the sagittal plane hip, knee and ankle angles because the majority of running-related injuries occur in the lower limbs. During the treadmill run, the kinematic data were recorded for a large number of consecutive strides for each individual (see Figure \ref{fig:sample-adjacent-strides-sub-01}). 
They were then segmented into individual strides, as a single stride is considered the most basic unit of analysis.
In human movement biomechanics, when multiple strides are available for each individual, they are typically reduced to a single ``representative stride" for analysis, which is usually an average (e.g., see a recent article by \textcite{fox_measurement_2023} discussing how many strides should be used to calculate the average). 
In \textcite{gunning_analyzing_2023}, we modelled the average hip and knee angle curves bilaterally for each subject using functional mixed effects modelling techniques.
However, collapsing the full collection of strides to a single summary curve is wasteful, as it discards information about stride-to-stride variability, serial autocorrelation among adjacent strides and changes in movement patterns over the course of the measurement period. Therefore, in this work we use LFDA to model repeatedly observed functional data in human movement biomechanics, allowing us to fully harness the rich dataset collected during the treadmill run.

Our motivating dataset has added complexities, which, when combined, require a novel modelling approach. 
Firstly, we want to employ a multivariate approach to capture the dependence among multiple joints (i.e., the hip, knee and ankle angles), rather than performing separate univariate analyses for each location. Multivariate functional models can be more efficient from a statistical perspective because strength is borrowed across the locations \parencite{zhu_multivariate_2017, volkmann_multifamm_2021}.
From an applied perspective, understanding the dependence (or co-ordination) among multiple joints is crucial for fully describing movement patterns \parencite{glazier_beyond_2021}.
Secondly, the participants were measured on both sides of the body, which adds a hierarchical structure to the data. Finally, we need to include scalar covariate information in our model, e.g., sex, running speed and injury status. This motivates the development of a \emph{multivariate multilevel longitudinal functional model}. 
The dataset contains more than \num{40000} multivariate functional observations from $284$ unique individuals, meaning it is large compared to typical datasets in biomechanics and other fields where FDA is routinely applied. 
This characteristic makes the computational feasibility of our proposed approach an important, additional consideration.
To the best of our knowledge, this is the first piece of work to develop statistical methodology to appropriately analyse repeatedly observed multivariate kinematic data in human movement biomechanics. While our motivating dataset comes from a short treadmill run, the methodology is applicable to various other settings in human movement biomechanics (e.g., longer running or walking sessions) and to data that are collected in other fields such as manufacturing and imaging.

The remainder of the article is structured as follows. In Section \ref{sec:related}, we summarise the existing literature on longitudinal functional models. In Section \ref{sec:model}, we describe our proposed methodology and its implementation. Section \ref{sec:simulation} contains a simulation study to illustrate the properties of the method under realistic data-generating scenarios. Section \ref{sec:results} contains the data analysis and results of our scientific application. We close with a discussion in Section \ref{sec:discussion}.

\section{Literature Review} \label{sec:related}

A variety of methods have been developed to model repeated functional observations from multiple individuals, e.g., \emph{functional multilevel} (or \emph{mixed effects} or \emph{hierarchical}) \emph{models} \parencite{morris_wavelet-based_2006, di_multilevel_2009, scheipl_functional_2015}. In this section, we focus on models that explicitly account for dependence in the repeated functional observations along a longitudinal timescale. Many of these approaches use a functional principal component analysis (FPCA) decomposition along the functional, or both functional and longitudinal, timescales.

\textcite{greven_longitudinal_2010} introduced longitudinal FPCA as an extension of multilevel FPCA \parencite[ml-FPCA;][]{di_multilevel_2009}, to capture linear longitudinal trends in longitudinal functional data. The ml-FPCA model consists of a subject-specific and curve-specific functional random intercept, each represented by a parsimonious FPCA decomposition. \textcite{greven_longitudinal_2010} extended this model to include a subject-specific functional random slope, which admits a joint FPCA decomposition with the subject-specific functional random intercept. To allow flexible non-parametric, rather than linear, longitudinal trends, \textcite{chen_modeling_2012} proposed a two-stage FPCA. In the first stage, they performed FPCA (on the functional timescale) separately at a grid of longitudinal time points. At the second stage, the first-stage functional principal component (FPC) scores were treated as functions of longitudinal time and subjected to a second FPCA decomposition. To model electroencephalogram (EEG) waveforms collected for multiple subjects over multiple trials at multiple electrodes located in different scalp regions, \textcite{hasenstab_multi-dimensional_2017} proposed a multilevel longitudinal FPCA, generalising the two-stage FPCA procedure of \textcite{chen_modeling_2012} to account for the different layers of variability in the longitudinal functional data (i.e., electrode within region within subject). 

\textcite{park_longitudinal_2015} proposed a more parsimonious two-stage FPCA, making the simplifying assumption that a \emph{longitudinal-time-invariant} FPCA basis can be used to represent the functions, with longitudinal trends captured only through the FPC scores. Thus, they performed a single ``marginal" FPCA at the first stage, ignoring the longitudinal time, and then treated each first-stage FPC score as a longitudinal time-varying function, decomposing it using a second-stage FPCA. The marginal approach alleviates the need to perform a separate first-stage FPCA at each longitudinal time point and was shown to be less computationally demanding than the ``conditional" approach of \textcite{chen_modeling_2012}. \textcite{li_fixed-effects_2022} recently developed hypothesis tests to choose the longitudinal covariance structure in the second stage of the method of \textcite{park_longitudinal_2015}, which allows the second-stage FPCA to be tested against a simpler parametric model (e.g., random intercept and slope). They also showed how the chosen covariance structure can be used to re-estimate a full functional mixed model including fixed effects of scalar covariates to improve estimation and inference. \textcite{chen_modelling_2017} proposed decompositions of longitudinal functional data based on marginal covariance structures, and introduced product FPCA to represent longitudinal functional observations on a tensor product basis of the marginal FPCs in the longitudinal and functional directions. \textcite{scheffler_hybrid_2020} extended product FPCA to account for dependence among longitudinally observed EEG functions from multiple regions, by treating the region as a dimension -- a discrete analogue of the longitudinal and functional dimensions. \textcite{lee_bayesian_2019} demonstrated how the very general basis modelling framework for functional mixed models, first proposed by \textcite{morris_wavelet-based_2006}, can incorporate longitudinally varying functions. They represented each function using a wavelet basis and modelled the basis coefficients separately using Bayesian scalar mixed effects models. The scalar mixed effects models included linear and smooth effects of scalar covariates and a small number of parametric basis functions as random effects to capture the longitudinal trends and account for them in fixed-effects estimation. \textcite{shamshoian_bayesian_2022} considered the product FPCA model from a Bayesian perspective, using a tensor-product basis representation of the longitudinal functions and a Bayesian latent factor model for the basis coefficients. Most recently, \textcite{boland_study_2022} constructed a dual time-frequency representation of EEG data from multiple trials, resulting in surfaces (functions of both time and frequency) that vary longitudinally (i.e., across trials). They performed a marginal multidimensional FPCA of the surfaces, ignoring longitudinal time, and then modelled longitudinal trends in the multidimensional FPCA scores using mixed effects models, with a small number of unpenalised B-spline basis functions used to capture smooth longitudinal trends.

All of the approaches described above have been developed for \emph{univariate} longitudinal functional data. In this work, we develop methodology for \emph{multivariate} (or \emph{vector-valued}) longitudinal functional data, to model multiple functional variables (i.e., the hip, knee and ankle angles) that vary longitudinally and have two nested levels of variability (side within subject). Although our approach is motivated by multiple kinematic variables, it could be extended to multivariate functional data with heterogeneous components \parencite[e.g., curves and images;][]{happ_multivariate_2018} varying longitudinally that might arise in other areas of research.

\section{Methodology}\label{sec:model}
\subsection{Model}
\begin{figure}[h]
    \centering
    \includegraphics[width = 0.9\textwidth, page = 2]{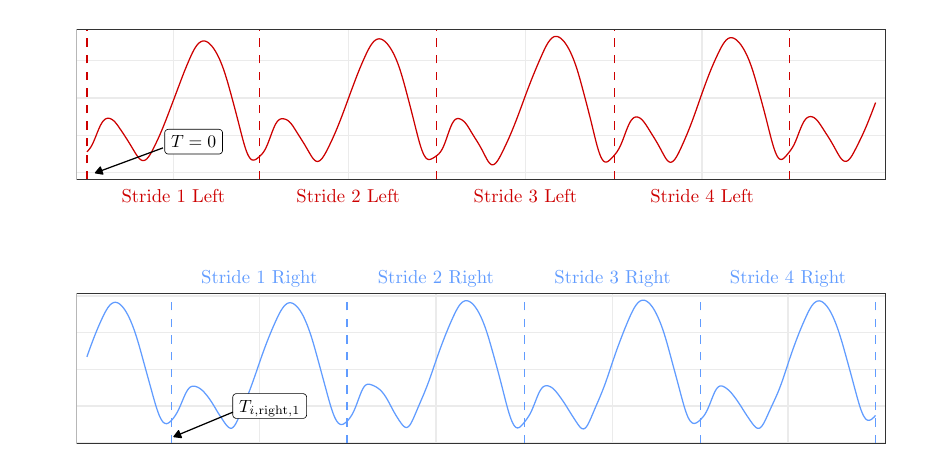}
    \caption{The timing of strides from the left and right sides of the body, illustrated using the sagittal knee angle functions. The top panel displays the left side sagittal knee angle curves. The bottom panel displays the right side sagittal knee angle curves.}
    \label{fig:stride-timings}
\end{figure}

We denote the multivariate functional observation from the $l$th stride for the $i$th individual on side $j$ as
$$
\boldy_{ijl} (t) = \left(y_{ijl}^{(hip)} (t), y_{ijl}^{(knee)} (t), y_{ijl}^{(ankle)} (t)\right)^\top,  \quad  l = 1, \dots n_{ij}, \quad j \in \{\text{left, right}\} \text{ and } i = 1, \dots, N, 
$$
where $N$ is the total number of individuals, $n_{ij}$ is the total number of strides taken by individual $i$ on side $j$, and $t \in [0, \ 100]$ is a normalised \emph{functional time} interval with $0$ representing the start of a stride and $100 (\%)$ representing the end. We also introduce a \emph{longitudinal time} variable $T$, such that $T_{ijl}$ indexes the time in the treadmill run at which stride $l$ occurs on side $j$ for subject $i$. The longitudinal time variable $T$ is also normalised so that $T \in [0, \ 1]$, where $0$ represents the start of the treadmill run and $1$ represents the end. 
The ordering and timing of strides is illustrated graphically in Figure \ref{fig:stride-timings}.
Finally, we let $\mathbf{x}_{ij} = (x_{ij1}, \dots, x_{ijA})^\top$ denote the vector of length $A$ of scalar covariates for subject $i$ on side $j$. The covariates could be subject specific (e.g., sex, height) or subject-and-side specific (e.g., an indicator for a subject's dominant side). However, we assume that the covariates are fixed across strides and hence $\mathbf{x}_{ij}$ is not indexed by $l$.

Our proposed multivariate multilevel longitudinal functional model is
$$
\boldy_{ijl} (t) = \boldbeta_0(t, T_{ijl}) + \sum_{a=1}^A x_{ija} \boldbeta_{a} (t) + \mathbf{u}_{i} (t, T_{ijl}) + \mathbf{v}_{ij} (t, T_{ijl}) + \boldsymbol{\varepsilon}_{ijl} (t).
$$
where $\boldbeta_0(t, T_{ijl})$ is the multivariate intercept function which varies smoothly in both functional and longitudinal time, $\boldbeta_{a} (t)$ is the multivariate functional fixed effect corresponding to the $a$th scalar covariate, $\mathbf{u}_{i} (t, T_{ijl})$ is the subject-specific multivariate functional random intercept that varies smoothly in both functional and longitudinal time, $\mathbf{v}_{ij} (t, T_{ijl})$ is the subject and side-specific multivariate functional random intercept that also varies smoothly in both functional and longitudinal time, and $\boldsymbol{\varepsilon}_{ijl} (t)$ is the smooth multivariate functional random error that is specific to observation $\boldy_{ijl} (t)$.

The intercept function $\boldbeta_0 (t, T)$ is assumed to be a smooth bivariate function of both functional time $t$ and longitudinal time $T$. Parametric models in the longitudinal direction are often assumed, such as constant $\boldbeta_0 (t, T) = \boldbeta_0(t)$ or linear $\boldbeta_0 (t, T) = \mathbf{b}_0 (t) + \mathbf{b}_1 (t) T$ \parencite{koner_second-generation_2023}. As described in Section \ref{sec:basis-representation}, we employ a more flexible approach, expanding $\boldbeta_0 (t, T)$ on a small number of parametric basis functions in the longitudinal direction. For $a = 1, \dots, A$, the multivariate functional fixed effect $\boldbeta_a (t)$ captures the influence of the $a$th scalar covariate on the ``expected level and shape" of the multivariate functional response \parencite{bauer_introduction_2018}. We assume that the multivariate functional fixed effects are constant across $T$, which implies that the scalar covariates affect the average running kinematics, rather than the kinematics at any particular point in the treadmill run. For $i = 1, \dots, N$, the subject-specific multivariate functional random intercept $\mathbf{u}_{i} (t, T)$ captures correlation among observations from the same subject. These functions are assumed to be independent realisations of a mean-zero multivariate Gaussian process with matrix-valued covariance function $\mathbf{Q}(t, t',T, T')$. Likewise, the subject-and-side-specific multivariate functional random intercepts $\mathbf{v}_{ij} (t, T)$ are assumed to be independent realisations of a mean-zero multivariate Gaussian process with matrix-valued covariance function $\mathbf{R}(t, t',T, T')$. These functions capture correlation among observations from the same subject and side. Finally, the multivariate functional random errors (or ``curve-level random effects") are assumed to be independent realisations of a zero-mean multivariate Gaussian process with matrix-valued covariance function $\mathbf{S}(t, t')$. The multivariate functional random error represents the deviation that is specific to observation $\boldy_{ijl} (t)$, i.e., what is not captured by the longitudinally varying subject-specific and subject-and-side-specific deviations. It is further assumed that the processes $\mathbf{u}_{i} (t, T)$, $\mathbf{v}_{ij} (t, T)$ and $\boldsymbol{\varepsilon}_{ijl}(t)$ are mutually uncorrelated.

\begin{figure}
    \centering
    \includegraphics[width = 1\textwidth]{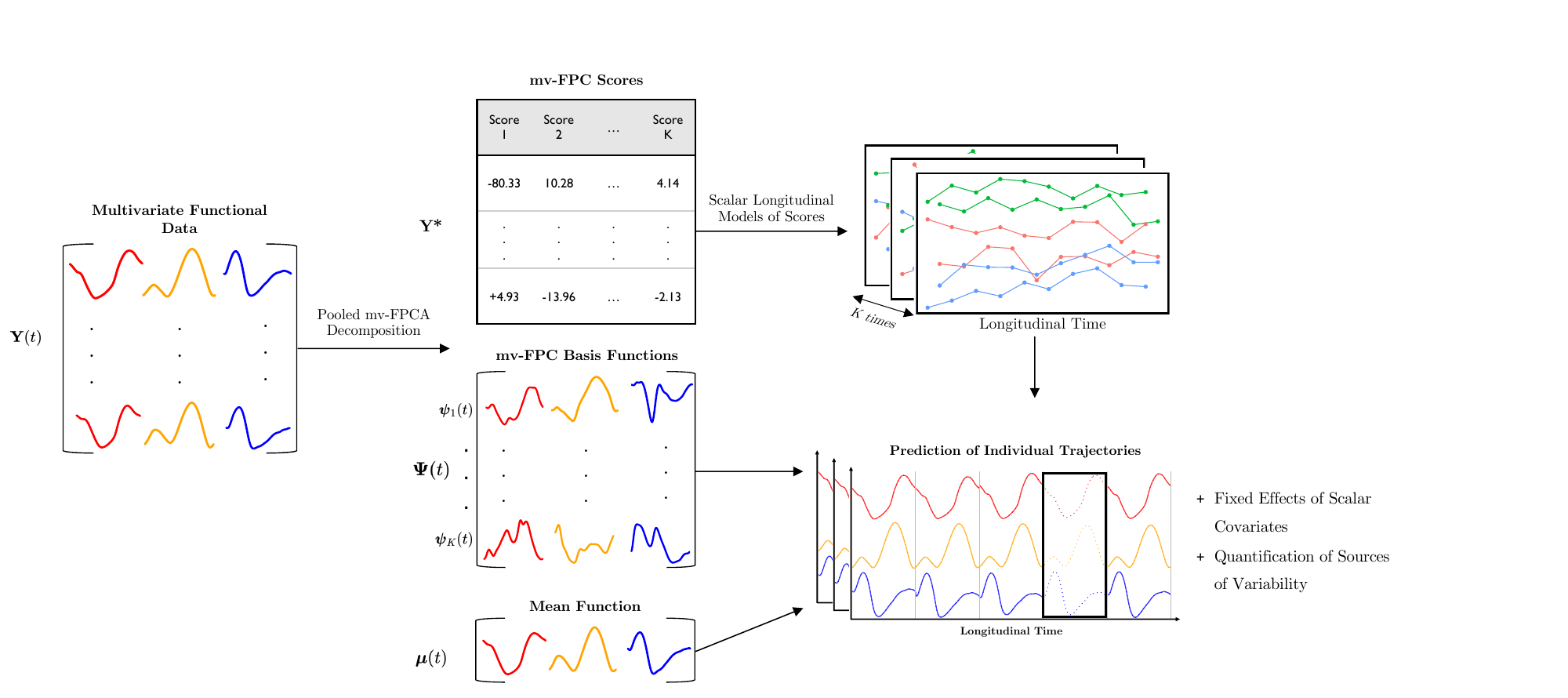}
    \caption{A flowchart of our approach to estimating the proposed multivariate multilevel longitudinal functional model.}
    \label{fig:method-flowchart}
\end{figure}

Our proposed methodology for estimating the model is described in the following sections and is summarised by the flowchart in Figure \ref{fig:method-flowchart}.

\subsection{Basis Representation of the Multivariate Functions} \label{sec:basis-representation}
For what follows, we assume that $\boldy_{ijl}(t)$ are centered, i.e., the overall functional sample mean $\widehat{\boldsymbol{\mu}}(t)$ has been subtracted from each observation. Our modelling approach mirrors many of the existing approaches for \emph{univariate} LFDA \parencite[e.g.,][]{park_longitudinal_2015, lee_bayesian_2019, boland_study_2022}, which is to first represent each multivariate functional observation by a basis expansion
$$
\boldy_{ijl}(t) = \sum_{k=1}^K y_{ijl, k}^* \boldpsi_k (t).
$$
The basis functions $\{\boldpsi_k(t)\}_{k=1}^{K}$ are multivariate functions and $y_{ijl, k}^*$ are scalar basis coefficients that weight the basis functions to produce the functional observations. We then model the scalar basis coefficients to capture longitudinal trends and the fixed effects of scalar covariates. For univariate functional data, the set of basis functions $\{\boldpsi_k(t)\}_{k=1}^{K}$ can be either known a priori \parencite[e.g., wavelets;][]{lee_bayesian_2019} or estimated from the data \parencite[e.g., FPCs;][]{aston_linguistic_2010}. For multivariate functional data, multivariate functional principal components (mv-FPCs) are a suitable choice, because they capture common variation among the dimensions of the multivariate function.

We calculate the mv-FPCs from the entire sample, ignoring the longitudinal and multilevel dependence structures. Using this pooled (or marginal) basis to represent all functions simplifies modelling and is necessary given the size and structure of the data at hand. Specifically, each basis function $\boldpsi_k (t)$ is a solution of the sample multivariate functional eigenequation
$$
\int_0^{100} \widehat{\mathbf{C}} (t, t') \boldpsi_k (t') \mathrm{d}t' = \lambda_k \boldpsi_k (t), \quad t \in [0, 100],
$$
where $\widehat{\mathbf{C}} (t, t')$ is the pooled matrix-valued covariance function 
$$
\widehat{\mathbf{C}} (t, t') = \frac{1}{N_{Total} - 1}\sum_{i=1}^N \sum_{j \in \{\text{left}, \text{right}\}} \sum_{l=1}^{n_{ij}} \boldy_{ijl}(t) \boldy_{ijl}(t')^\top,
\quad t, t' \in [0, 100],
$$
where $N_{Total} =\sum_{i=1}^N \sum_{j \in \{\text{left}, \text{right}\}} n_{ij}$ is the total number of observations. 
We compute the mv-FPCs in a two-stage approach, by first expanding the observations within each dimension on a univariate B-spline basis. In the second stage, the basis coefficients from the different dimensions are combined into a single matrix, and  classical multivariate principal component analysis (PCA) is performed on a weighted version of this combined matrix of B-spline basis coefficients, where the weights are given by the inner product between the spline basis functions; the full calculation is given in \textcite{jacques_model-based_2014, happ_multivariate_2018}. A choice remains regarding the value of $K$, the number of mv-FPCs to retain. The eigenvalue $\lambda_k$ represents the amount of variance explained by the $k$th mv-FPC. The eigenvalues typically decay rapidly, meaning that a small number of mv-FPCs explain a large proportion of the variance. In this work, we opt for a \emph{near-lossless} basis representation, which retains almost all of the information in the observed multivariate functional data \parencite{morris_comparison_2017, lee_bayesian_2019}. 
This allows the basis coefficients to be treated as transformed data rather than estimated parameters and modelled in place of the observed multivariate functions, as described in Section \ref{sec:modelling-basis-coefs} \parencite{morris_automated_2011}.
We thus choose $K$ such that a high percentage (e.g., $99.5\%$) of the variance in the data is explained.
For a chosen $K$, we perform a ten-fold cross-validation procedure, in which the data from each subject are included in only one fold to avoid data leakage, to estimate the (overall) out-of-sample variance explained.
Additionally, we perform leave-one-subject-out cross-validation to estimate the average percentage of variance explained \emph{within} each subject.

\subsection{Modelling the Basis Coefficients}\label{sec:modelling-basis-coefs}

We model the $N_{Total} \times K$ matrix $\mathbf{Y}^*$ of basis coefficients (i.e., mv-FPC scores) in place of the observed multivariate functional data. We make the simplifying assumption that each of the $K$ basis coefficients (i.e., each column of $\mathbf{Y}^*$) can be modelled separately \parencite{morris_wavelet-based_2006, aston_linguistic_2010, park_functional_2017, boland_study_2022}. Although this assumption may not be flexible enough to fully capture the dependence in the individual random processes \parencite{koner_second-generation_2023}, we have shown empirically in \textcite{gunning_analyzing_2023} that it works well for a simpler model of this type. Importantly, the assumption simplifies the problem to fitting a separate univariate scalar longitudinal model to each basis coefficient.

\begin{figure}[h]
    \centering
    \includegraphics[page = 4, width = 0.9\textwidth]{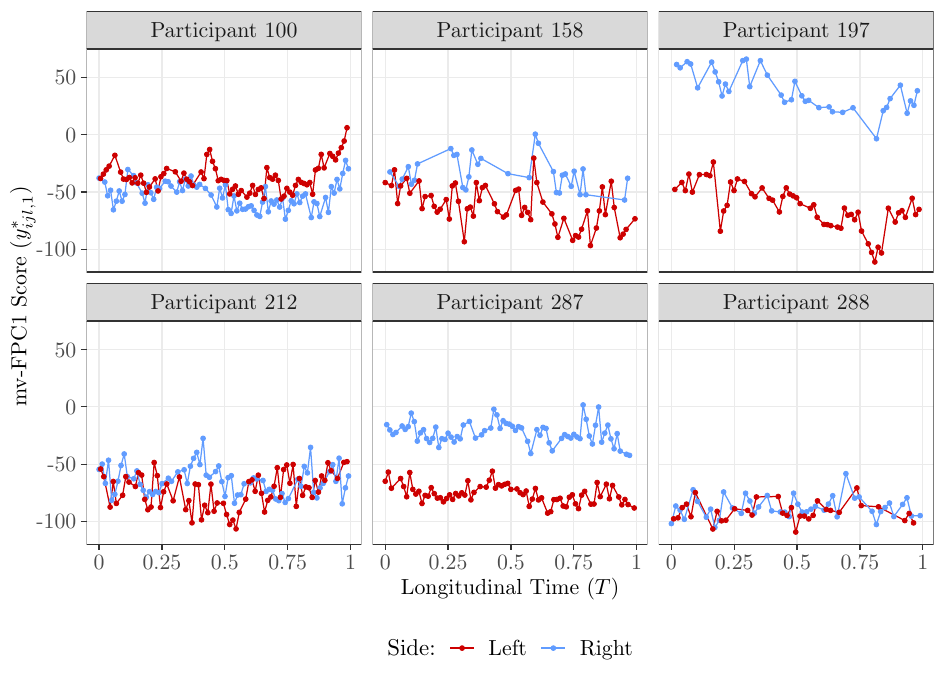}
    \caption{Longitudinal trajectories of the first mv-FPC score for a random sample of six subjects.}
    \label{fig:mv-fpc1-score}
\end{figure}

The model for the $k$th basis coefficient is
\begin{equation} \label{eq:univariate-scalar-model}
    y^*_{ijl, k} = \beta_{0, k}^* (T_{ijl}) + \sum_{a=1}^A x_{ia} \beta_{a, k}^*  + u_{i, k}^* (T_{ijl}) + v_{ij, k}^* (T_{ijl}) + \varepsilon_{ijl, k}^*,
\end{equation}
which is a multilevel functional model in longitudinal time $T$ \parencite{di_multilevel_2009}. Figure \ref{fig:mv-fpc1-score} displays the longitudinal trajectories of the first basis coefficient (i.e., the first mv-FPC score, labelled mv-FPC1) for six randomly selected subjects.
We choose to parameterise the longitudinally varying functions using a small number of unpenalised basis functions. 
This basis is chosen because we expect changes in the longitudinal direction to be smooth and simple -- treadmill running is a stable form of movement, especially as the participants ran at a fixed speed for the duration of the run.
For ease of presentation, we use the same set of basis functions $\{\xi_d(T) \}_{d=1}^D$ to represent each longitudinally varying term, giving
$$
\beta_{0, k}^* (T) = \sum_{d=1}^D \beta_{0, k, d}^* \ \xi_d(T), \quad  u_{i, k}^* (T) = \sum_{d=1}^D u_{i, k, d}^* \ \xi_d (T) \quad \text{and} \quad v_{ij, k}^* (T) = \sum_{d=1}^D v_{ij, k, d}^* \ \xi_d (T).
$$
However, a reduced (or different) set of basis functions can be used to represent any of the three terms, and a different basis can be used for each $k$.
The choice of basis $\{\xi_d(T) \}_{d=1}^D$ still remains. 
\textcite{lee_bayesian_2019} found, via an exploratory analysis, that the longitudinal trends in their wavelet basis coefficients were well modelled by a hyperbolic basis, whereas \textcite{boland_study_2022} used a constant function and a small number of B-spline basis functions.
In this work, we use a small number of natural cubic B-spline basis functions to represent each term. We also investigate the use of a separate ml-FPCA basis estimated directly from the data for each $k$.

Substituting the basis function evaluations into model \eqref{eq:univariate-scalar-model} gives, for the $k$th basis coefficient, the model
$$
 y^*_{ijl, k} = \sum_{d=1}^D \beta_{0, k, d}^* \ \xi_d (T_{ijl}) + \sum_{a=1}^A x_{ia} \beta_{a, k}^*  + \sum_{d=1}^D u_{i, k, d}^* \ \xi_d (T_{ijl}) +\sum_{d=1}^D v_{ij, k, d}^* \ \xi_d (T_{ijl}) + \varepsilon_{ijl, k}^*,
$$
where $(u_{i, k, 1}^*, \dots, u_{i, k, D}^*)^\top \sim \mathcal{N}(\mathbf{0}, \mathbf{Q}^*_k)$, $(v_{ij, k, 1}^*, \dots, v_{ij, k, D}^*)^\top \sim \mathcal{N}(\mathbf{0}, \mathbf{R}^*_k)$ and $\varepsilon_{ijl, k}^* \sim \mathcal{N}(0, s_k)$. This is a scalar linear mixed effects model \parencite{laird_random-effects_1982}, and can be fitted using any standard software, e.g., the \pkg{lme4} \proglang{R} package \parencite{bates_fitting_2015}. The matrices $\mathbf{Q}^*_k$ and $\mathbf{R}^*_k$ are of dimension $D \times D$ and contain $D(D+1) / 2$ free parameters to estimate. To reduce computational overhead and the problems that arise when estimating large unstructured covariance matrices in mixed effects models \parencite[e.g.,][]{bates_parsimonious_2018}, \textcite{lee_bayesian_2019} and \textcite{boland_study_2022} both made the assumption that these matrices are diagonal.
% We only make this assumption when it is justified by, e.g., the use of a ml-FPCA basis so it can be seen as a special case.
In general, we do not make this assumption except when it is justified by the basis functions being used (e.g., when using a ml-FPCA basis).
The scalar linear mixed effects models are fitted separately for each $k$ using Restricted Maximum Likelihood (REML).

\subsection{Reconstructing the Model Terms}
\subsubsection{Fixed Effects}

Rather than inspect individual parameter estimates, it is more natural to combine the estimated parameters across the basis coefficients to reconstruct and estimate the functional model terms. The estimated intercept function is given by
$$
\widehat{\boldbeta}_0 (t, T) = \sum_{k=1}^K \sum_{d=1}^D \widehat{\beta}^*_{0, k, d} \ \xi_d (T) \boldsymbol{\psi}_k (t),
$$
where $\widehat{\beta}^*_{0, k, d}$ denotes the estimate of $\beta^*_{0, k, d}$ from the mixed effects model. Likewise, the estimate of the functional fixed effect of the $a$th scalar covariate is given by
$$
\widehat{\boldbeta}_a(t) = \sum_{k=1}^K \widehat{\beta}_{a, k}^* \boldsymbol{\psi}_k (t), \quad a~=~1\dots, A.
$$
The estimates of $\widehat{\Var}(\widehat{\beta}_{a, k}^*)$ from the mixed effects model can be combined across $k$ to construct approximate pointwise and simultaneous confidence bands for $\boldbeta_a(t)$, as described in \textcite{gunning_analyzing_2023}. We also use a non-parametric bootstrap procedure, resampling subjects with replacement, to quantify uncertainty in the estimated parameters \parencite[e.g.,][]{crainiceanu_bootstrap-based_2012, park_simple_2018, cui_fast_2022}.

\subsubsection{Covariance Structures}

The matrix-valued covariance function $\mathbf{Q}(t, t', T, T')$ implied by the model is
$$
    \mathbf{Q}(t, t', T, T') =  \Expec[\mathbf{u}_i(t, T) \ \mathbf{u}_i(t', T')^\top] = \boldsymbol{\Psi} (t)^\top (\mathbb{I}_K \otimes \boldsymbol{\xi} (T))^\top \mathbf{Q}^* (\mathbb{I}_K \otimes \boldsymbol{\xi} (T')) \boldsymbol{\Psi} (t'),
$$
where $\boldsymbol{\Psi}(t)$ is the $K \times 3$ matrix containing the mv-FPCs, $\boldsymbol{\xi} (T) = (\xi_1(T), \dots, \xi_D(T))^\top$ and $\mathbf{Q}^*$
is the block-diagonal matrix containing the matrices $\mathbf{Q}^*_1, \dots, \mathbf{Q}^*_K$ along its diagonal. Similarly, we have that
$$
\mathbf{R}(t, t', T, T') =  \Expec[\mathbf{v}_{ij}(t, T) \ \mathbf{v}_{ij}(t', T')^\top] = \boldsymbol{\Psi} (t)^\top (\mathbb{I}_K \otimes \boldsymbol{\xi} (T))^\top \mathbf{R}^* (\mathbb{I}_K \otimes \boldsymbol{\xi} (T')) \boldsymbol{\Psi} (t'),
$$
where $\mathbf{R}^*$ is the block-diagonal matrix containing the matrices $\mathbf{R}^*_1, \dots, \mathbf{R}^*_K$ along its diagonal. Finally, the within-function covariance is 
$$
\mathbf{S}(t, t') = \boldsymbol{\Psi} (t)^\top \mathbf{S}^* \boldsymbol{\Psi} (t'), \quad \mathbf{S^*} = \diag\{s_1, \dots, s_K\}.
$$

\subsubsection{Individual Trajectories}

Our methodology facilitates the prediction of subject-specific and subject-and-side-specific trajectories at any point in the treadmill run. The prediction of the subject-specific multivariate functional random intercept at any $T \in [0, 1]$ is given by
$$
\widehat{\mathbf{u}}_i (t, T) = \sum_{k=1}^K \sum_{d=1}^D \widehat{u}_{i, k, d}^* \ \xi_d (T) \boldpsi_k (t), \quad i = 1, \dots, N,
$$
where $\widehat{u}_{i, k, d}^*$ is the Best Linear Unbiased Predictor (BLUP) of $u_{i, k, d}^*$ from the linear mixed effects model. The subject-and-side specific deviation is obtained analogously as
$$
\widehat{\mathbf{u}}_i (t, T) + \widehat{\mathbf{v}}_{ij} (t, T) = \sum_{k=1}^K \sum_{d=1}^D \bigl( \widehat{u}_{i, k, d}^* + \widehat{v}_{ij, k, d}^* \bigr) \ \xi_d (T) \boldpsi_k (t), \quad i = 1, \dots, N, \quad \text{and} \quad j \in \{\text{left}, \text{right} \}.
$$
The predicted trajectories can be used, for example, to investigate change in technique over the course of the treadmill run as measured by the rate of change with respect to $T$. To assess the predictions, we create a ``test set" by holding out ten strides per subject and side at random points throughout the treadmill run. These observations are left out for both the mv-FPC computation and model fitting. We use the test set to visualise the model's predictions of held-out strides. We also compare test-set prediction error to that of a ``naive" model that ignores the longitudinal dependence structure \parencite{park_longitudinal_2015}. We could alternatively remove only the final functional observation for each individual, as proposed by \textcite{park_longitudinal_2015}, whose goal was to forecast future disease progression as measured by their functional observations. However, as data-collection errors in motion capture (e.g., marker problems) often mean that certain strides have to be removed at different points in the treadmill run, it is valuable to understand how well the model can impute the missing strides given the data at other points.

\section{Simulation} \label{sec:simulation}

This section presents a simulation study to assess the properties of the proposed methodology under data-generating scenarios that resemble our application.
We are interested in the computational efficiency, the quality of estimated model parameters and accuracy of predictions of individual observations.
We investigate these measures while varying the number of subjects, number of observations per subject and the longitudinal dependence structure in the multivariate functional observations. 

\subsection{Simulation Setup}
 
We use a basis expansion to generate longitudinal multivariate functional observations. We use the first 10 empirical mv-FPCs, which explain $95\%$ of the variance in the data application in Section \ref{sec:results}, as basis functions and then generate observations by simulating the basis coefficients from scalar multilevel longitudinal models. Specifically, we generate observations as 
\begin{equation}\label{eq:data-generating-model}
    \mathbf{y}_{ijl} (t) = \widehat{\boldsymbol{\mu}} (t) + \sum_{k=1}^{10} y_{ijl, k}^* \widehat{\boldsymbol{\Psi}}_k(t), \quad  l = 1, \dots n_{ij}, \quad i = 1, \dots, N, \quad \text{and} \quad j \in \{\text{left, right}\},
\end{equation}
where $\widehat{\boldsymbol{\mu}} (t)$ and $\{\widehat{\boldsymbol{\Psi}}_k(t)\}_{k=1}^{10}$ are the empirical mean and mv-FPCs from the application in Section \ref{sec:results}. Each basis coefficient is generated according to the scalar multilevel longitudinal model
$$
y^*_{ijl, k} = \sum_{d=1}^D \beta_{0, k, d}^* \ \xi_d (T_{ijl}) + \sum_{a=1}^2 x_{ia} \beta_{a, k}^*   + \sum_{d=1}^D u_{i, k, d}^* \ \xi_d (T_{ijl}) + \sum_{d=1}^D v_{ij, k, d}^* \ \xi_d (T_{ijl}) + \varepsilon_{ijl, k}^*,
$$
where $(u_{i, k, 1}^*, \dots, u_{i, k, D}^*)^\top \sim \mathcal{N}(\mathbf{0}, \mathbf{Q}^*_k)$, $(v_{ij, k, 1}^*, \dots, v_{ij, k, D}^*)^\top \sim \mathcal{N}(\mathbf{0}, \mathbf{R}^*_k)$ and $\varepsilon_{ijl, k}^* \sim \mathcal{N}(0, s_k)$. The subject-specific scalar covariates for sex and age, denoted by $x_{i1}$ and $x_{i2}$, are drawn from binomial and Gaussian distributions, respectively, to mimic the self-selected running speed and sex covariates in our application. The empirical effect estimates for these covariates are used for $\beta_{a,k}^*$, $a=1,2$ and $k = 1, \dots, 10$. Orthogonal versions of the polynomial basis functions $\xi_1 (T) = 1$,  $\xi_2 (T) = T$, and $\xi_3 (T) = T^2$ are used for the longitudinally varying terms, and empirical estimates for $\mathbf{Q}^*_k$, $\mathbf{R}^*_k$, $s_k$ and $\beta_{0, k, d}^*$, $k = 1, \dots, 10$ and $d = 1, \dots, 3$ are based on an initial model fit using this basis. Observations are generated at $n_{ij} = 80$ equally-spaced points on $[0, 1]$ for each subject but, as described below, not all observations are included in the final model fit. Additional details on the simulation setup are provided in Appendix \ref{sec:appendix-simulation-settings}.

\begin{figure}
    \centering
    \includegraphics[width = 0.75\textwidth]{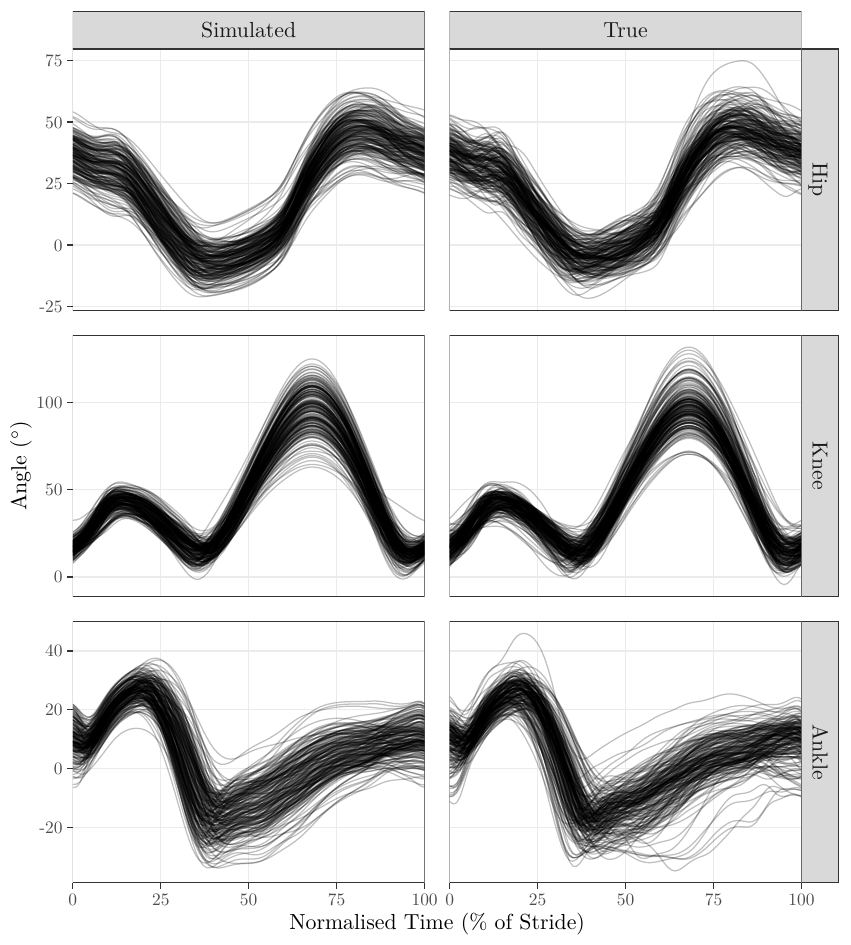}
    \caption{A sample of $200$ observations randomly sampled from a simulated dataset under the baseline simulation scenario (left) and from the true dataset (right). }
    \label{fig:simulated-vs-true}
\end{figure}

The following parameters are varied one at a time from their baseline (first) level:
\begin{enumerate}
    \item Number of subjects: $N = 280$, $N = 500$ and $N = 1000$. 
    \item Proportion of missing strides: $0.1$, $0.2$ and $0.5$. 
    \item Strength of the longitudinal variation: $1$, $2$ and $3$.
\end{enumerate}
The number of subjects is varied to understand improvements in performance and the increase in computational overhead as the sample size is increased. The proportion of missing strides is varied because subjects in our dataset have differing numbers of strides, with some removed because of, e.g., data-collection errors. A baseline proportion of $0.1$ are removed to facilitate the construction of a test set to evaluate model predictions. Our reasoning for increasing the strength of the longitudinal variation is that the longitudinal trends observed in our application are small relative to the constant between-subject variability. Therefore, we increase the longitudinal variation by suitably rescaling the elements of $\mathbf{Q}^*_k$ and $\mathbf{R}^*_k$ to double and triple the contributions of the non-constant basis functions $\xi_2 (T)$ and $\xi_3 (T)$. Figure \ref{fig:simulated-vs-true} displays $200$ randomly-sampled observations from a simulated dataset under the baseline simulation scenario (left) and from the true dataset (right). Marginally, at least, the generative model appears to produce functional observations that resemble the real data.

% In total, there are seven simulation scenarios (the baseline scenario plus a scenario in which each of the three parameters to their two levels). 
% For each scenario, we perform $500$ simulation replications.
% A variance-explained threshold of $99.5\%$ is used to choose the number of mv-FPCs to retain.

In the simulation, we fit four models that parameterise the longitudinally varying random effects differently.
We refer to them as the polynomial, naive, spline and ml-FPCA models.
The polynomial model is correctly specified in that it employs the polynomials used to generate the data as longitudinal basis functions.
The naive model ignores longitudinal variation in the random effects and employs just a random intercept at both the subject and subject-and-side levels.
The spline model uses three natural cubic spline basis functions (plus a constant function) as longitudinal basis functions.
We tried using this basis to represent the random effects at both the subject and subject-and-side levels.
However, as discussed in the real data analysis in Section \ref{sec:modelling-results}, the majority of model fits in an initial simulation were singular.
Therefore, we simplify the spline model by dropping the longitudinally varying basis at the highest (subject-and-side) level.
The ml-FPCA model uses longitudinal basis functions at both levels that are estimated directly from the data.
For the naive, spline and ml-FPCA models, the longitudinally varying intercept is represented using three natural cubic spline basis functions.
In each of the seven simulation scenarios, we perform $500$ simulation replications.
A variance explained threshold of $99.5\%$ is used to choose the number of mv-FPCs to retain in each replicate.

\subsection{Evaluation Criteria}\label{sec:simulation-evaluation}

Performance is evaluated in the different scenarios using a number of criteria. We record the computation time for both the mv-FPCA step and the modelling of the basis coefficients. Each fixed effect estimate is evaluated in terms of integrated squared error (ISE). Letting $\mathcal{P} = \{\text{hip}, \text{ knee}, \text{ ankle} \}$, the ISE for the intercept function is
$$
\text{ISE}(\widehat{\boldsymbol{\beta}}_0) = \frac{1}{100} \sum_{p \in \mathcal{P}} \int_0^{100} \int_0^{1} \{\widehat{\beta}^{(p)}_0 (t, T) - \beta^{(p)}_0 (t, T) \bigr \}^2 \mathrm{d}T \mathrm{d}t,
$$
and for the effects of the scalar covariates is
$$
\text{ISE}(\widehat{\boldsymbol{\beta}}_a) = \frac{1}{100} \sum_{p \in \mathcal{P}} \int_0^{100} \{ \widehat{\beta}^{(p)}_a (t) - \beta^{(p)}_a (t) \bigr\}^2 \mathrm{d}t, \quad a = 1, 2.
$$
Prediction of held-out observations is evaluated in terms of integrated squared prediction error (ISPE). If observation $\mathbf{y}_{ijl}(t)$ is included in the test set, then
$$
\text{ISPE}(\widehat{\boldsymbol{y}}_{ijl}) = \frac{1}{100} \sum_{p \in \mathcal{P}} \int_0^{100} \{ \widehat{y}_{ijl}^{(p)} (t) - y_{ijl}^{(p)} (t) \bigr \}^2 \mathrm{d}t,
$$
where
$$
\widehat{\boldsymbol{y}}_{ijl} (t) = \boldsymbol{\widehat{\beta}}_0 (t, T_{ijl}) + \sum_{a=1}^2 x_{ia} \widehat{\boldbeta}_a (t) + \widehat{\mathbf{u}}_i (t, T_{ijl}) + \widehat{\mathbf{v}}_{ij} (t, T_{ijl}).
$$
On each simulation replicate, the average ISPE over all of the test-set observations is recorded.
We do not evaluate the estimated mv-FPCs in this section but a note on their estimation is provided in Appendix \ref{app:eigenfunction-recovery}.

\subsection{Simulation Results}

\begin{figure}
    \centering
    \includegraphics[width = 1\textwidth]{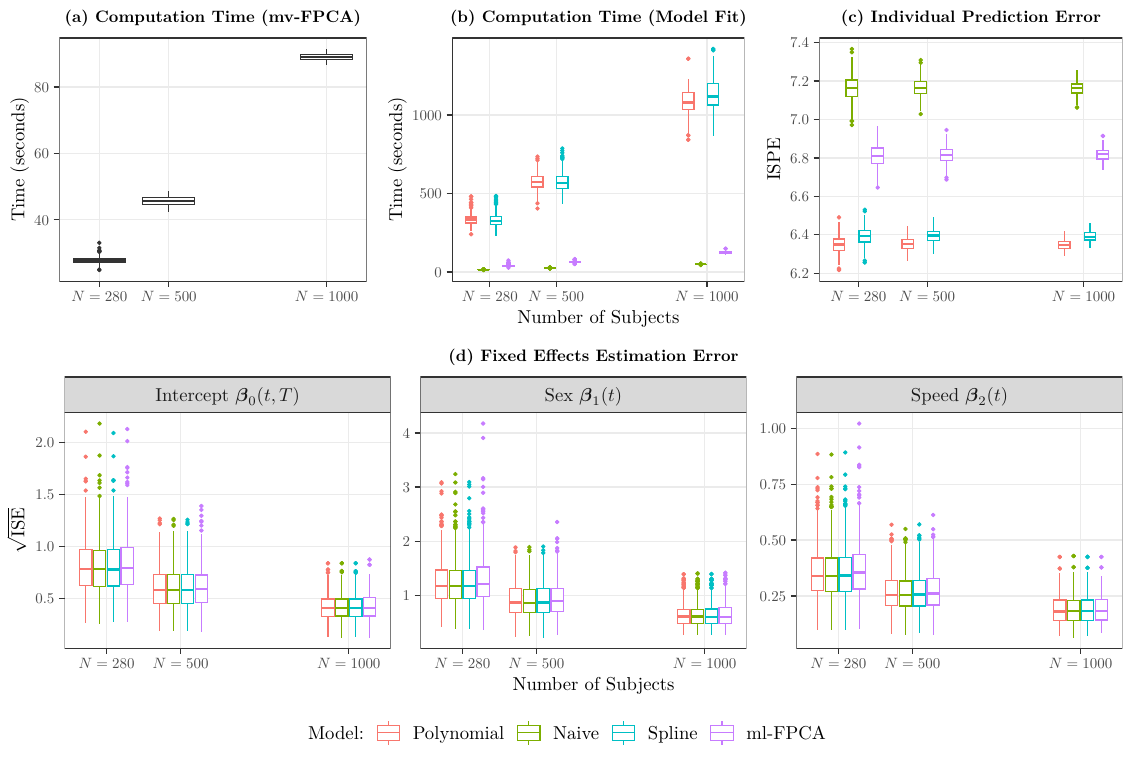}
    \caption{Results of the simulation varying the number of subjects, $N$. \textbf{(a)} The computation time for the mv-FPCA step in seconds. \textbf{(b)} The computation time for the model fits in seconds.
    \textbf{(c)} The integrated squared prediction error of held-out strides from the test set.
    \textbf{(d)} The integrated squared error of the fixed effects estimates.
    The strength of the longitudinal variation and the proportion of missing strides are fixed at their baseline values of $1$ and $0.1$, respectively.}
    \label{fig:simulation-sample-size}
\end{figure}

In this section, we present the results of varying the number of subjects $N$ between the levels $280$ (baseline), $500$ and $1000$ (Figure \ref{fig:simulation-sample-size}).
As expected, the computation time of both the mv-FPCA step (Figure \ref{fig:simulation-sample-size} (a)) and of each of the model fits (Figure \ref{fig:simulation-sample-size} (b)) increases with $N$.
For all three values of $N$, the naive model has the lowest computational effort because it estimates the fewest parameters.
The ml-FPCA model is the next fastest computationally because a parsimonious longitudinal basis is being used and the random effects covariance matrices $\mathbf{Q}_k^*$ and $\mathbf{R}_k^*$ are assumed to be diagonal, meaning fewer parameters are being estimated. The polynomial and spline models are comparable in terms of computation time.
The prediction error of individual observations appears to remain stable across the three values of $N$ (Figure \ref{fig:simulation-sample-size} (c)). 
Predictably, the naive model has the worst prediction accuracy (largest ISPE) and the correctly-specified polynomial model has the best prediction accuracy (smallest ISPE).
The ISPE of the spline model is reasonably close to that of the polynomial model indicating that the spline is approximating the longitudinally varying terms at the subject level well and that the amount of longitudinal variation that it ignores at the subject-and-side level is small.
The ml-FPCA model's ISPE is higher but still better than that of the naive model -- a higher variance explained cutoff might be needed to estimate a ml-FPCA basis as flexible as the spline or polynomial bases.
Figure \ref{fig:simulation-sample-size} (d) displays the results of the estimation of the three fixed effects parameters. The error in the fixed effects estimates is similar for all four models and, as anticipated, improves as $N$ increases.

Full results of varying the proportion of missing strides and the strength of longitudinal variation are included in Appendix \ref{app:extra-simulation-results}.
To summarise, computation time decreased and individual prediction error increased (most substantially for the ml-FPCA model) as the proportion of missing observations was increased. Increasing the strength of the longitudinal variation resulted in increased improvements in prediction error of the polynomial, spline and ml-FPCA models over the naive model.
Fixed effects estimation remained relatively unchanged across the scenarios.

\section{Data Analysis and Results} \label{sec:results}

\subsection{Data Collection, Extraction and Preparation}

This paragraph is a synopsis of the experimental setup, data collection and extraction process, with full details provided in the thesis of \textcite[pp. 180-183]{dillon_investigation_2022}.
Recreational runners aged between $18$ and $64$ years of age with no history of injury in the last three months were recruited as participants for the RISC study.
Prior to the baseline testing session, in which the kinematic data were collected, the participants completed an online survey regarding their injury history, training history and demographics.
To prepare for the testing, participants completed a dynamic lower-body warm-up routine and a 6-minute treadmill run (Runner-DTM2500, Flow Fitness, Amsterdam, Netherlands) to familiarise themselves with the treadmill.
Then, they ran for three minutes at a self-selected speed that represented their typical training pace, while kinematic data were collected using a 17-camera, three-dimensional motion analysis system (Vantage, Vicon, Oxford, United Kingdom) for the first full minute of the run. 
The motion data (i.e., marker trajectories) were sampled at a rate of \si{200\hertz} and filtered using a fourth-order zero-lag Butterworth filter at \SI{15}{\hertz} to smooth out observational errors. 
From the filtered trajectories, the sagittal plane hip, knee and ankle angles were extracted bilaterally for the first minute of the treadmill run based on the Vicon Plug in Gait model \parencite{vicon_plug-gait_2022} and the ``OSSCA" method for functional joints in Vicon Nexus 2 \parencite{taylor_repeatability_2010}.

The long sequences of kinematic measurements (e.g., Figure \ref{fig:sample-adjacent-strides-sub-01}) were segmented into individual strides based on the initial contact of the foot with the ground, which was identified using a custom algorithm. The univariate functional data for each stride were time normalised and registered to the point of the maximum knee flexion angle, which is a clear and easily identifiable landmark in each stride. Within each dimension, $80$ cubic B-spline basis functions were used to provide a near-lossless representation of the univariate functions. For each stride, the longitudinal time variable $T$ was created based on the time at which that stride started, with $T = 0$ representing the start of the subject's capture period. This variable was  normalised by dividing by the subject's maximum capture time, so that $T \in [0, 1]$. 
Subject-specific normalisation has been criticised from an interpretability perspective \parencite{park_longitudinal_2015}, but it is reasonable in our case as, although some subjects were recorded for longer or shorter than 1 minute, the average capture period was exactly 1 minute and the majority ($93\%$) of subjects' capture periods were between $50$ and $70$ seconds.
The test set was constructed by randomly selecting $10$ observations (i.e., the multivariate functional data from $10$ strides) separately on the right and left side for every subject. To achieve a minimum of $10$ strides on each side for every subject in both the training and testing sets, four subjects who had fewer than $20$ strides on either side were excluded from the analysis. In total, the dataset used in the analysis consisted of \num{47150} multivariate functional observations from $284$ subjects, with \num{41470} included in the training set and \num{5680} in the test set. Table \ref{tab:tab1.} contains summary characteristics of the dataset.

\begin{table}
\centering
\begin{tabular}[t]{llrr}
\toprule
  &    & \textbf{Mean} & \textbf{Std. Dev.}\\
\midrule
Speed (\si{\km \per \hour}) &  & 11.0 & 1.6\\
Age (years) &  & 43.3 & 9.0\\
Weight (kg) &  & 72.3 & 12.9\\
Height (cm) &  & 172.8 & 9.7\\
\midrule
 &  & \textbf{N} & $\mathbf{\mathbf{(\%)}}$\\
\midrule
Retrospective Injury Status & Never Injured & 48 & 16.9\\
 & Injured $>2$ yr. ago & 66 & 23.2\\
 & Injured $1-2$ yr. ago & 51 & 18.0\\
 & Injured $<1$ yr. ago & 119 & 41.9\\
Sex & Male & 173 & 60.9\\
 & Female & 111 & 39.1\\
\bottomrule
\end{tabular}
\caption{Summary characteristics of the dataset used in the analysis.}
\label{tab:tab1.}
\end{table}

The mv-FPCA, computed from the univariate basis expansions, yielded $K=27$ mv-FPCs to explain $99.5\%$ of the variance in the multivariate functional data. Ten-fold cross-validation estimated the overall percentage of variance explained at approximately $99.5\%$ and leave-one-subject-out cross-validation estimated the average percentage of variance explained \emph{within} each subject at $95.7\%$.

\subsection{Modelling Results}\label{sec:modelling-results}
As in \textcite{gunning_analyzing_2023}, all of the subject-specific covariates in Table \ref{tab:tab1.} were included as fixed effects in the model. A constant function and four natural cubic B-splines were used as longitudinal basis functions, with unstructured $\mathbf{Q}_k^*$ and $\mathbf{R}_k^*$ matrices. After inspecting initial models that converged to a singular fit, we dropped the longitudinally varying basis at the subject-and-side level and retained the constant function (i.e., random intercept) at this level, giving the following simplified model
\begin{align} 
    \mathbf{y}_{ijl}(t) =  \  &\boldbeta_0(t, T_{ijl}) + \sum_{a=1}^3 x_{ia}  \boldbeta_a (t) + \  \text{speed}_i \times \boldbeta_4 (t) + \text{sex}_i \times \boldbeta_5 (t) +  \text{age}_i \times   \boldbeta_6 (t) \\ &+  \text{weight}_i \times \boldbeta_7 (t) + \text{height}_i \times \boldbeta_8 (t) + \mathbf{u}_i (t, T_{ijl}) + \mathbf{v}_{ij}(t) + \boldsymbol{\varepsilon}_{ijl}(t),
\end{align}
where $x_{i1}, x_{i2}$ and $x_{i3}$ are dummy-coded variables representing the ``Injured more than 2 years ago", ``Injured 1-2 years ago" and ``Injured less than 1 year ago" categories of the retrospective injury status variable, where the reference category is ``Never injured", $\text{speed}_i$ is the self-selected running speed of subject $i$ in \si{\km \per \hour}, $\text{sex}_i$ is a dummy-coded variable for the sex of subject $i$ ($0=$ male, $1=$ female), $\text{age}_i$ is the age of subject $i$ in years, $\text{weight}_i$ is the weight of subject $i$ in kilograms and $\text{height}_{i}$ is the height of subject $i$ in centimetres. All numeric variables were centred to make the intercept function more interpretable.

A naive model which fixed $\mathbf{u}_i(t, T) = \mathbf{u}_i(t)$ and a model using an empirically-determined ml-FPCA longitudinal basis were also used for comparison of the fitted trajectories and test-set predictions. For each mv-FPC score, fixed effects were estimated under a working independence assumption and the fast ml-FPCA method \parencite{cui_fast_2023} was used to estimate a ml-FPCA longitudinal basis that explained $99.5\%$ of the variability at both levels. The estimated ml-FPCA basis functions were then used to re-fit the model, with diagonal $\mathbf{Q}_k^*$ and $\mathbf{R}_k^*$ matrices \parencite{li_fixed-effects_2022, leroux_fast_2023}. 

The computation times for fitting the spline, ml-FPCA and naive models were $10.5$ minutes, $3.2$ minutes and $0.5$ minutes, respectively, on a $2019$ MacBook Pro with 8 GB of memory. The non-parametric bootstrap for the spline model was performed in parallel across $7$ cores and took $12.84$ hours to complete.
% The fitted trajectories from these models are used in the comparisons in Section \ref{sec:ranef_result}.

\subsubsection{Fixed Effects}
\begin{figure}
    \centering
    \includegraphics[width = 0.8\textwidth]{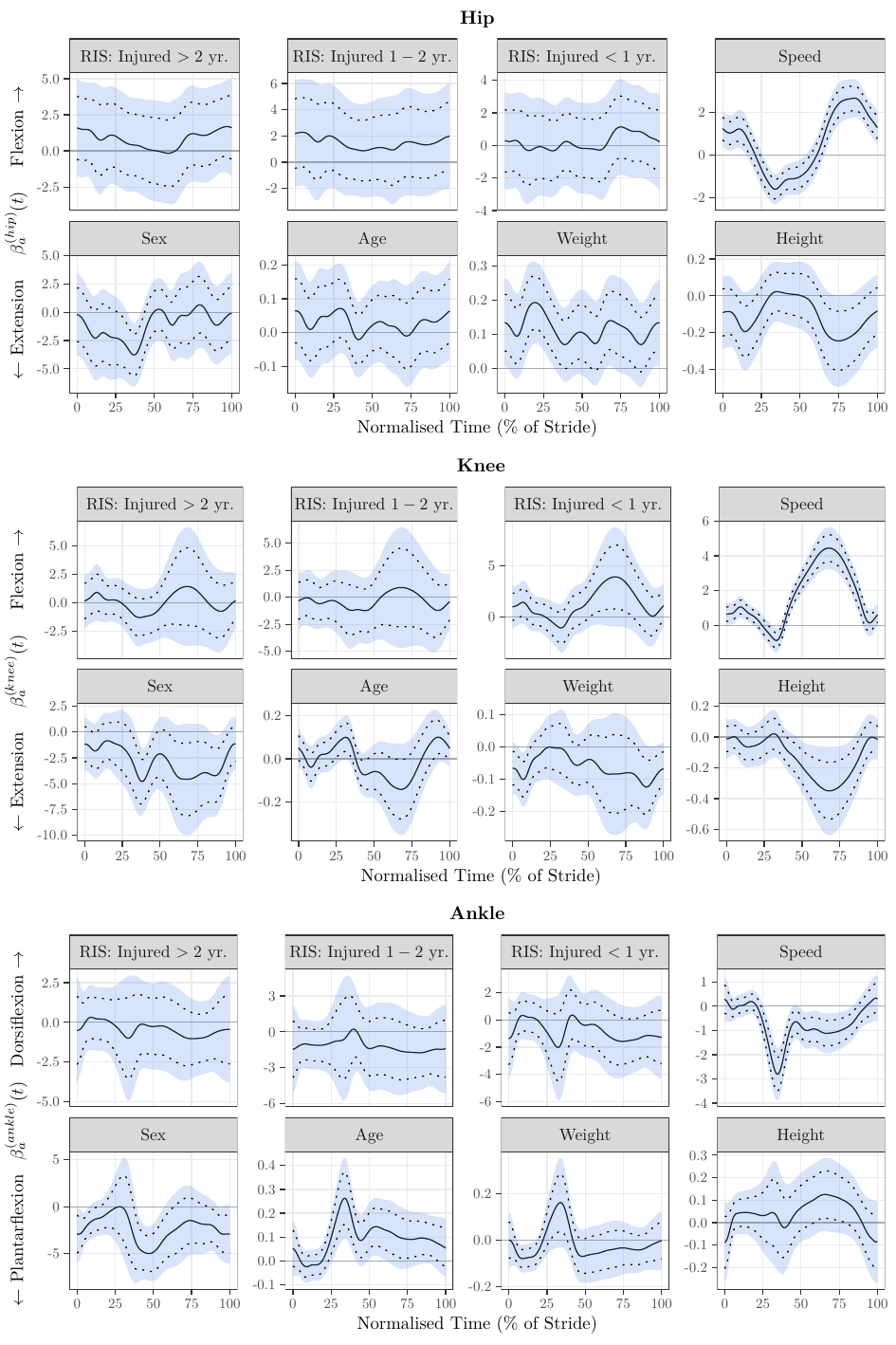}
    \caption{The estimated coefficient functions of the fixed effects from the fitted model. The black solid line represents the point estimate, the dotted black lines indicate pointwise $95\%$ confidence intervals and the light blue ribbons represent $95\%$ simultaneous confidence bands.}
    \label{fig:fixef-coef-plot}
\end{figure}
% The fixed effects structure of our model was
% \begin{align} 
%     \boldbeta_0(t, T_{ijl}) + \sum_{a=1}^3 x_{ia}  \boldbeta_a (t) + \  \text{speed}_i \times \boldbeta_4 (t) + \text{sex}_i \times \boldbeta_5 (t) +  \text{age}_i \times   \boldbeta_6 (t) &+  \text{weight}_i \times \boldbeta_7 (t) \\ &+  \ \text{height}_i \times \boldbeta_8 (t),
% \end{align}
% where $x_{1}, x_{2}$ and $x_{3}$ are dummy-coded variables representing the ``Injured more 2 years ago", ``Injured 1-2 years ago" and ``Injured less than 1 year ago" categories of the retrospective injury status variable, where the reference category is ``Never injured", $\text{speed}_i$ is the self-selected running speed of subject $i$ in {\km \per \hour}, $\text{sex}_i$ is a dummy-coded variable for sex of subject $i$ ($0=$ male, $1=$ female), $\text{age}_i$ is the age of subject $i$ in years, $\text{weight}_i$ is the weight of subject $i$ in kilograms and $\text{height}_{i}$ is the height of subject $i$ in centimetres.

Analysis of the functional coefficients of the longitudinal basis functions used to model the intercept revealed that it was approximately constant in the longitudinal direction (Appendix \ref{app:fixed-effects}). 
Figure \ref{fig:fixef-coef-plot} displays the estimated coefficient functions that capture the effects of scalar covariates in our model.
The solid black line represents the point estimate, the dotted black line represents a $95\%$ pointwise confidence interval and the light blue ribbons represent $95\%$ simultaneous confidence bands.
In all three dimensions, the simultaneous confidence bands for the retrospective injury status coefficient functions contain zero (solid grey horizontal line) for all $t$, indicating that there is no evidence of a significant difference between any of the categories and the reference category of ``Never injured".
We observe a strong, noticeable effect of self-selected running speed in all three dimensions, as the simultaneous confidence bands only contain zero around the time that the point estimate crosses $0$.
Running at a higher speed is associated with greater hip flexion at initial contact and late in the swing phase ($t>60\%$) and greater hip extension around the time of toe-off ($t\approx38\%$), greater knee flexion which is most pronounced in the stance phase around the time of peak knee flexion angle ($t \approx 69\%$) and increased ankle plantarflexion which is most pronounced around the time of maximum plantarflexion ($t\approx38\%$). 
These effects are consistent with those found in smaller biomechanical studies that employed more elementary statistical analyses (e.g., using discrete variables and treating speed as a fixed category) \parencite{orendurff_little_2018, fukuchi_public_2017}.
The coefficient functions for the effect of sex are large in magnitude, reaching almost $5^\circ$ in the knee and ankle. However, the corresponding confidence bands are wide and contain zero for almost all $t$, indicating a lot of uncertainty about this effect. There is limited evidence of an age, height or weight effect. Although the simultaneous confidence bands for these coefficient functions do not contain zero at certain points, the magnitude of each effect is small.
As expected, the fixed effects estimates for the hip and knee are almost identical to those presented in \textcite{gunning_analyzing_2023}, where the average (rather than individual) strides were analysed.

\subsubsection{Random Effects}\label{sec:ranef_result}

\begin{figure}[h]
    \centering
    \includegraphics[width = 0.95\textwidth]{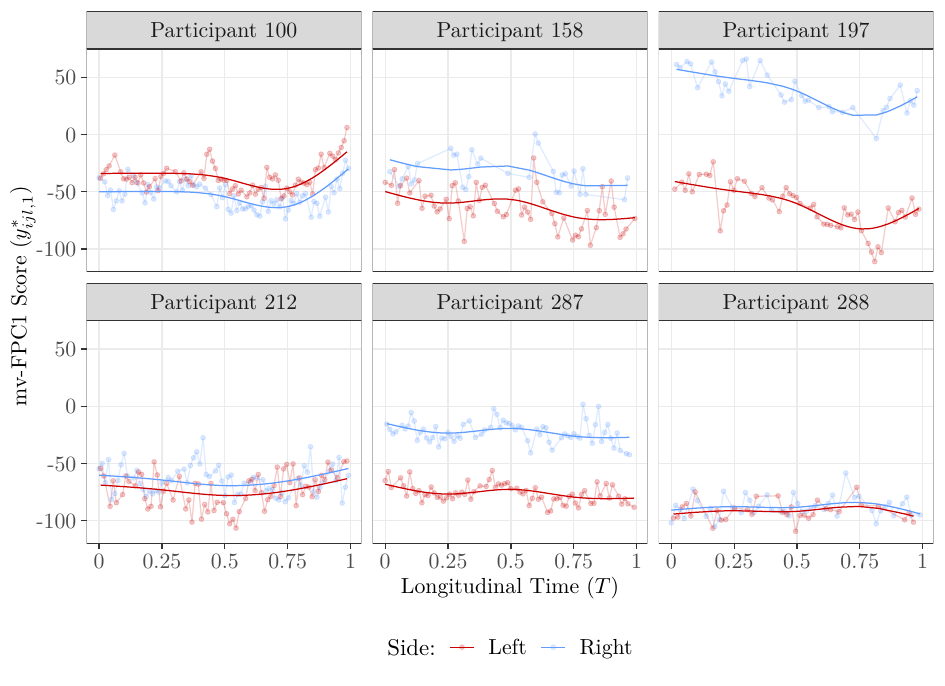}
    \caption{The longitudinal trajectories of the first mv-FPC score for the random sample of six subjects presented in Figure \ref{fig:mv-fpc1-score}. The fitted subject-and-side specific trajectories are overlaid as solid lines.}
    \label{fig:fitted-mv-fpc1-score}
\end{figure}

\begin{figure}
    \centering
    \includegraphics[width = 0.9\textwidth]{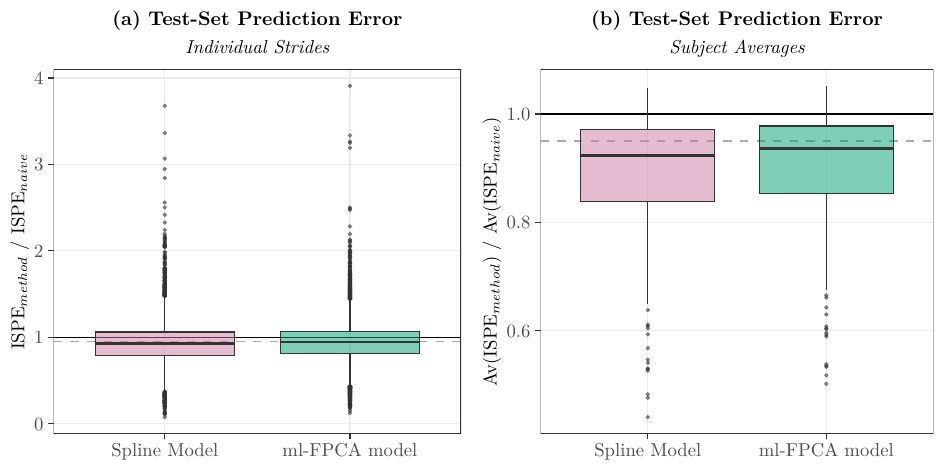}
    \caption{The prediction errors of held out strides from the test set for the spline model and the ml-FPCA model, presented relative to the naive model. $\mathbf{(a)}$ The ratio of the models' ISPE to the naive model's ISPE for individual strides in the test set. $\mathbf{(b)}$ The ratio of the models' average ISPE to the naive model's average ISPE for individual subjects in the test set. A grey horizontal dashed line is added at $0.95$ and a black solid line is at $1$.}
    \label{fig:test-set-PE}
\end{figure}

In this section, we present analysis of the fitted subject-and-side specific trajectories, which are obtained as BLUPs of the random effects. Figure \ref{fig:fitted-mv-fpc1-score} displays the same trajectories presented in Figure \ref{fig:mv-fpc1-score}, this time with the model fits overlaid.
This sample of trajectories is representative of the majority of subjects in the dataset -- they are relatively stable over the course of the treadmill run and exhibit only modest changes.
% Overall, the fitted trajectories for each score for the full dataset resembled the sample in Figure \ref{fig:fitted-mv-fpc1-score}, being relatively stable over the course of the treadmill run and exhibiting modest changes.
This is also reflected in the test-set prediction error, where the average ratio of the ISPE of the longitudinal model to the ISPE of the naive model was $0.93$, indicating that the longitudinal model provides a $7\%$ reduction in prediction error relative to the naive model (Figure \ref{fig:test-set-PE} (a)).
Figure \ref{fig:test-set-PE} (b) displays the ratio of the average ISPE of the longitudinal model to the average ISPE of the naive model for each subject. From this plot, it is evident that the longitudinal model provides a modest improvement over the naive model for almost every subject.
% From inspecting this ratio for the mean average ISPE per subject in Figure \ref{fig:test-set-PE} (b), it is evident that the longitudinal model provides a modest improvement over the naive model for almost every subject.
In both panels, there appears to be little difference in the improvements provided by our spline model and the ml-FPCA model.

\begin{figure}
    \centering
    \includegraphics[width = 1\textwidth]{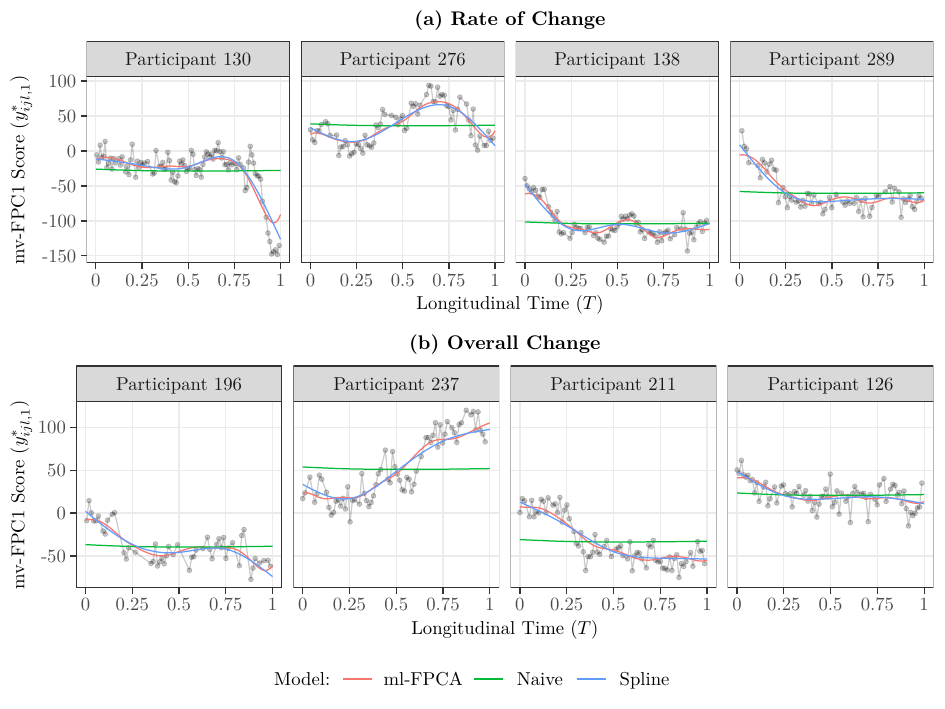}
    \caption{Observed and fitted values of the mv-FPC1 score for subjects identified based on summaries from the model. \textbf{(a)} The top four subjects based on the integrated squared first derivative of their fitted trajectory with respect to longitudinal time. \textbf{(b)} The top four subjects based on the overall change during the treadmill run. The dark grey dots and lines represent the observed data and coloured lines indicate the fitted trajectories from the ml-FPCA, naive and spline models. Only the left side data and fits for each subject are shown to avoid over-plotting. Two of the subjects in \textbf{(a)} also ranked in the top four for overall changes but were excluded from \textbf{(b)} to avoid duplication.}
    \label{fig:change-plot}
\end{figure}

Figure \ref{fig:change-plot} displays fits for subjects that were chosen according to summaries from the model. 
Firstly, we calculated the integrated squared first derivative with respect to longitudinal time of each subject's fitted profile, which provides a measure of the rate of change (or deviation from a constant fit) over the course of the treadmill run. 
Figure \ref{fig:change-plot} (a) displays the first mv-FPC score for the top four subjects ranked according to this metric. For ease of interpretation, we have only displayed the left side observations. All four subjects exhibit non-stationary patterns that are captured well by the longitudinal models (both spline and ml-FPCA). The naive model, which assumes that each individual's deviation is constant across longitudinal time, is inadequate.
Figure \ref{fig:change-plot} (b) displays another four subjects, this time ranked according to a simpler metric -- the overall change in the subject's fitted profile over the course of the run, calculated as the absolute difference between the subjects' fitted profiles at $T=0$ and $T=1$. 
Two subjects from Figure \ref{fig:change-plot} (a) also ranked in the top four for this metric but were excluded to avoid duplication in the figure. Non-stationary trends, which cannot be captured by the naive model, are evident again. It should be noted that these summaries were computed based on the full multivariate function but we have displayed the first mv-FPC score. However, this mv-FPC captured the largest amount of variance in the longitudinal direction, so it is a reasonable choice.

\begin{figure}
    \centering
    \includegraphics[width = 0.9\textwidth]{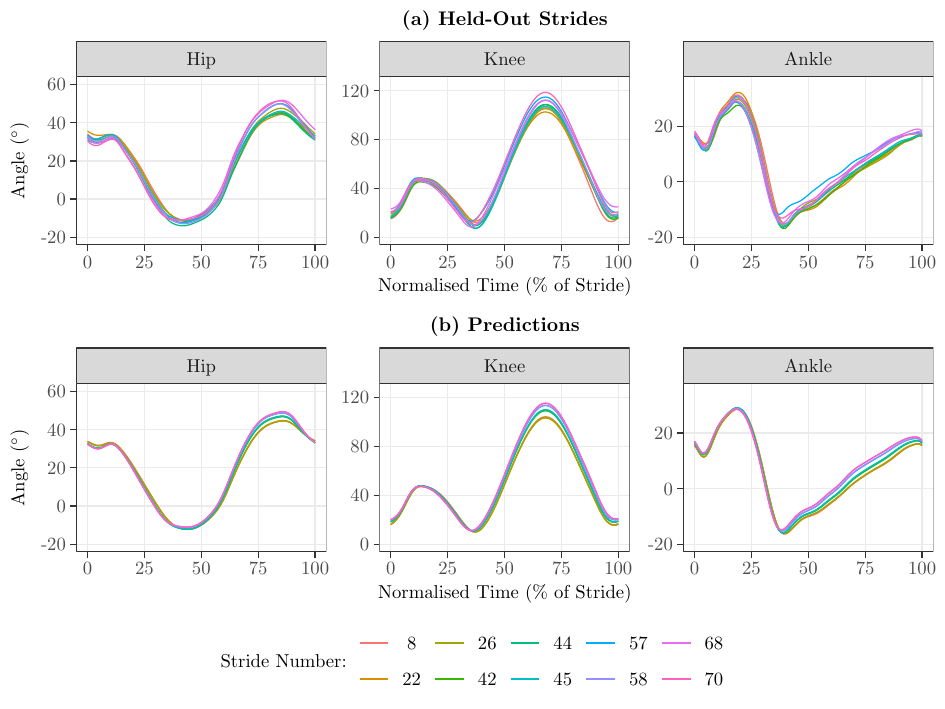}
    \includegraphics[width = 0.8\textwidth]{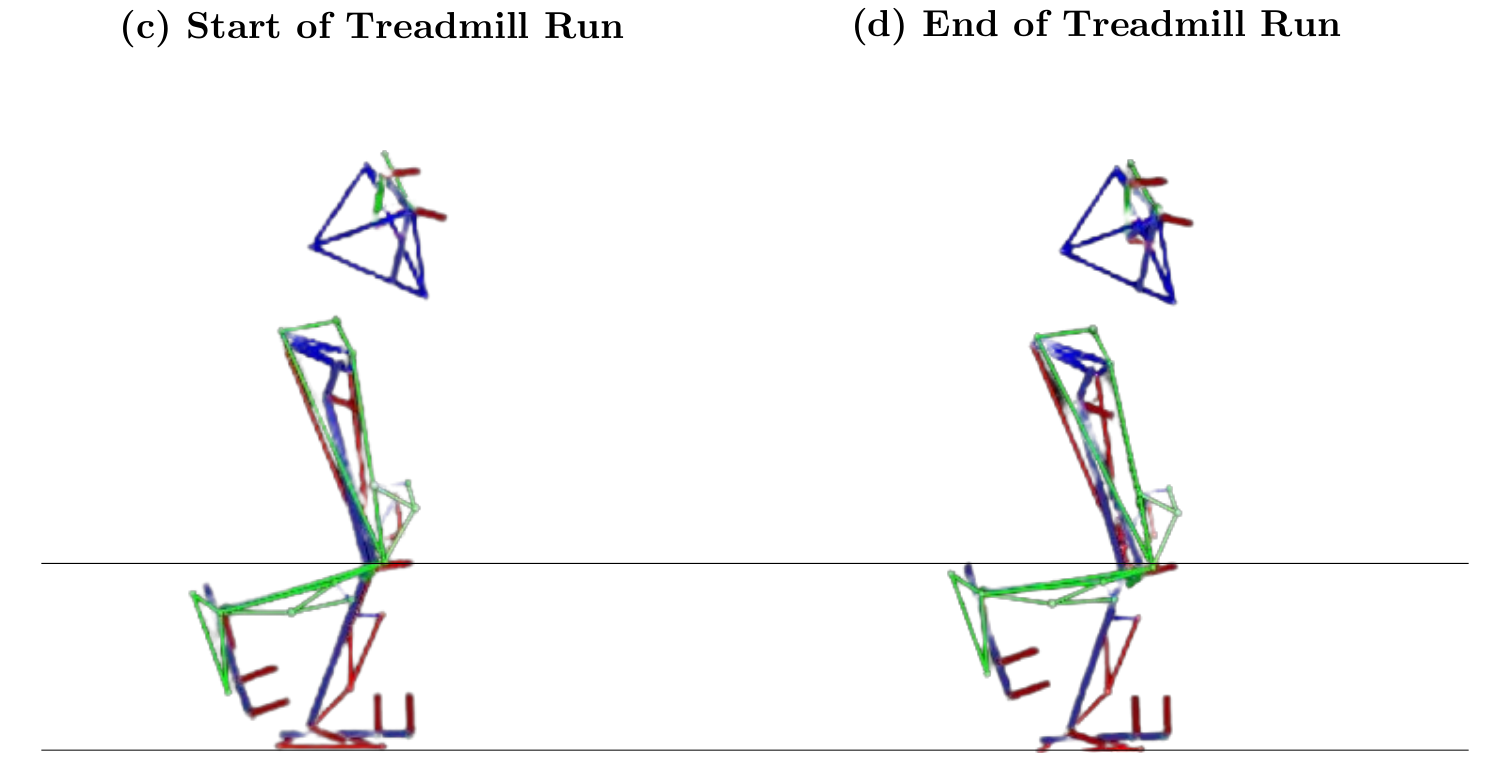}
    \caption{Individual analysis for Participant 237. \textbf{(a)} The held-out strides from the test set for this subject, coloured according to stride number. \textbf{(b)} Model predictions of the strides in \textbf{(a)}. \textbf{(c)} The motion-capture animation for this subject at the time of peak knee flexion angle at the start of the treadmill run (stride 1). \textbf{(d)} The motion-capture animation for this subject at the time of peak knee flexion angle at the end of the treadmill run (stride 80).}
    \label{fig:p4237}
\end{figure}

As the mv-FPC scores in Figure \ref{fig:change-plot} are a level of abstraction away from the multivariate functional data, we examine the fitted multivariate functions for a single individual. Based on Figure \ref{fig:change-plot} (b), we choose to display Participant 237 because they exhibited a consistent, almost-linear evolution.
Figure \ref{fig:p4237} (a) and (b) display this subject's held-out strides from the test set and predicted values from the model, respectively. They are displayed on a rainbow-style plot, where the colour of the line indicates the stride number and hence the longitudinal time \parencite{hyndman_rainbow_2010, shang_grouped_2017}. The trends in the held-out strides in the swing phase ($t>38\%$ of stride) are captured reasonably well in the model predictions (i.e., the colouring of the observed data and the predicted curves appears consistent). Figure \ref{fig:p4237} (c) and (d) display the motion-capture animation at the time of peak knee flexion angle for this subject at the start (stride 1) and end (stride 80) of the treadmill run, respectively. The difference in the two pictures reflects the changes across longitudinal time that are evident in Figure \ref{fig:p4237} (a) and (b), in particular the greater knee flexion at the end of the treadmill run.

\section{Discussion} \label{sec:discussion}

% \subsection{Summary}

We have presented a novel multilevel multivariate longitudinal functional model for kinematic data collected during a treadmill run. 
From a methodological perspective, this work has extended existing ideas for univariate longitudinal functional data \parencite[e.g.,][]{park_longitudinal_2015} to the multivariate (functional) case. 
From a computational perspective, our approach can be implemented using existing open-source software
and is computationally feasible for our motivating dataset, which contains almost \num{50 000} multivariate functional observations. 
In the application, the model allowed us to quantify and visualise the average effects of scalar covariates on the multivariate functional data, which were consistent with existing results in the biomechanical literature.
Unlike conventional approaches for analysing these data, our model appropriately accounts for the longitudinal dependence in the repeated strides, which allowed us to capture meaningful individual changes over the course of the short treadmill run.
A number of directions for future work, both methodological and applied, are discussed below.

% We modelled the longitudinal trends in the mv-FPC scores using a small number of unpenalised basis functions because we expected changes over the course of the treadmill run to be reasonably simple. 
% Alternatively, an emprically-determined ml-FPCA basis could be used to avoid assuming any structure a-priori. {\color{purple}We explore this alternative method in Appendix X, showing that it produces similar results for our application}. 
% We also allowed the fixed effects, subject-level and subject-and-side-level random effects to jointly enter the model through the mv-FPC scores to borrow strength in estimation and inference. 
% A potential drawback of this choice is that the same pooled mv-FPCA basis is used to represent all model terms.
% An alternative would be to estimate the fixed effects under a working independence assumption \parencite[e.g.,][]{cederbaum_functional_2016, di_multilevel_2009} and apply our method to the resulting residuals. Alternatives to using the marginal mv-FPCA basis for all model terms, such as using separate bases for $\mathbf{u}_i(t, T)$, $\mathbf{v}_{ij} (t, T)$ and $\boldsymbol{\varepsilon}_{ijl}(t)$, possibly based on ideas from product FPCA \parencite{chen_modelling_2017}, also present an interesting line of future work.

From an applied perspective, this work opens up a large number of exciting avenues to explore. The ability to model repeated movement data over the course of a long measurement period presents a variety of opportunities. For example, now that we have developed a model for the full collection of strides for each individual, we could use the fitted subject-specific longitudinal profiles to cluster individuals or to predict a scalar outcome (e.g., prospective injury). While we restricted the fixed effects of scalar covariates to be longitudinal time-invariant, we could imagine instances where smooth effects in both functional and longitudinal time would be of considerable interest. For instance, in addition to having different average movement patterns, we might expect different groups of individuals to fatigue differently and hence exhibit smooth differences in the longitudinal direction. 
Our approach could enable detailed investigations of whether consistency of running or walking technique is dependent on different running surfaces \parencite{mohr_whole-body_2023}, affected by different neurological conditions \parencite{pieruccini-faria_gait_2021} or can be maintained through biofeedback \parencite{argunsah_bayram_influence_2021}.
Due to the rise of wearable sensor technologies (e.g., inertial sensors), we expect an abundance of human movement data to be collected repeatedly for large numbers of individuals over the coming years both in one-off running sessions and also on multiple occasions on a run-by-run basis.
Although wearable-sensor data have different features to motion-capture data (e.g., sampling rates and signal-to-noise ratios) that may require modifications to pre-processing or modelling, the ideas presented in this work form the basis for building flexible, interpretable models for human movement analysis.

Investigating alternative methodological choices would also be of interest in future work. For example, instead of using basis functions to capture longitudinal dependence in the scalar linear mixed models, an AR(1) dependence structure for the errors could be trialled \parencite{chi_models_1989}. The type of functional covariance structure implied by an AR(1) model for the mv-FPCA scores could then be studied, similar to how \textcite{zhang_functional_2016} did for spatial autoregressive models. Likewise, while we adopted a basis modelling approach by projecting the data onto a common mv-FPCA basis and modelling the mv-FPC scores, it would be interesting to examine the use of different bases for different terms in the model. This might, however, be computationally demanding.

Finally, some limitations of our work are as follows. Firstly, we modelled the time-normalised and registered functional data. Although this is common in gait analysis because the start, end and landmark points are well defined and practically meaningful, doing so ignores the presence of phase variability. Future work should extend ideas from \textcite{hadjipantelis_unifying_2015}
to jointly model the phase parameters alongside the mv-FPC scores. Second, on  examining regression diagnostics of the scalar linear mixed models, we found that the conditional residual distributions were heavy tailed due to the presence of outliers (Appendix \ref{sec:lfmm-diagnostics}).
Although the linear mixed model has been shown to be very robust to non-Gaussian error distributions \parencite{jacqmin-gadda_robustness_2007, asar_linear_2020, knief_violating_2021}, future work could consider the use of robust FPCA and linear mixed models.
A final limitation is that our data-generating model does not guarantee continuity between adjacent strides (i.e., that the end of one stride and the start of the next stride match).
For the fitted subject-and-side specific trajectories, this is a very minor issue because changes in the longitudinal direction are smooth and any discontinuities are very minor.
For now, smoothing over any discontinuities when simulating individual strides would be a simple fix.
Future work could investigate how a suitable constraint could be put on the curve-specific smooth error to enforce continuity.

\section*{Acknowledgment}
This work was supported in part by Science Foundation Ireland (SFI) under grant numbers 18/CRT/6049 (EG), 19/FFP/7002 (SG, AJS and NB), and SFI/12/RC/2289\_P2 (RISC running dataset), and co-funded by the European Regional Development Fund.
The authors wish to acknowledge the Irish Centre for High-End Computing (ICHEC) for the provision of computational facilities and support.

% ---------------------------------------------------------------------------
% APPENDIX
% \bibliographystyle{abbrvnat}
% %\renewcommand*{\bibfont}{\small}
% \bibliography{references.bib}

\printbibliography

@article{pieruccini-faria_gait_2021,
	title = {Gait variability across neurodegenerative and cognitive disorders: {Results} from the {Canadian} {Consortium} of {Neurodegeneration} in {Aging} ({CCNA}) and the {Gait} and {Brain} {Study}},
	volume = {17},
	copyright = {© 2021 The Authors. Alzheimer's \& Dementia published by Wiley Periodicals LLC on behalf of Alzheimer's Association},
	issn = {1552-5279},
	shorttitle = {Gait variability across neurodegenerative and cognitive disorders},
	url = {https://onlinelibrary.wiley.com/doi/abs/10.1002/alz.12298},
	doi = {10.1002/alz.12298},
	language = {en},
	number = {8},
	urldate = {2023-09-08},
	journal = {Alzheimer's \& Dementia},
	author = {Pieruccini-Faria, Frederico and Black, Sandra E. and Masellis, Mario and Smith, Eric E. and Almeida, Quincy J. and Li, Karen Z. H. and Bherer, Louis and Camicioli, Richard and Montero-Odasso, Manuel},
	year = {2021},
	note = {\_eprint: https://onlinelibrary.wiley.com/doi/pdf/10.1002/alz.12298},
	pages = {1317--1328},
}

@article{argunsah_bayram_influence_2021,
	title = {The influence of biofeedback on physiological and kinematic variables of treadmill running},
	volume = {21},
	issn = {2474-8668},
	url = {https://doi.org/10.1080/24748668.2020.1861898},
	doi = {10.1080/24748668.2020.1861898},
	number = {1},
	urldate = {2023-09-08},
	journal = {International Journal of Performance Analysis in Sport},
	author = {Argunsah Bayram, Hande and Yalcin, Begum},
	month = jan,
	year = {2021},
	note = {Publisher: Routledge
\_eprint: https://doi.org/10.1080/24748668.2020.1861898},
	pages = {156--169},
}

@article{mohr_whole-body_2023,
	title = {Whole-body kinematic adaptations to running on an unstable, irregular, and compliant surface},
	volume = {(Advance Online Publication {https://doi.org/10.1080/1476} {3141.2023.2222022})},
	issn = {1476-3141},
	url = {https://doi.org/10.1080/14763141.2023.2222022},
	doi = {10.1080/14763141.2023.2222022},
	urldate = {2023-09-08},
	journal = {Sports Biomechanics},
	author = {Mohr, M. and Peer, L. and De Michiel, A. and van Andel, S. and Federolf, P.},
	month = jun,
	year = {2023},
	pmid = {37317805},
	note = {Publisher: Routledge
\_eprint: https://doi.org/10.1080/14763141.2023.2222022},
}

@article{scheipl_functional_2015,
	title = {Functional {Additive} {Mixed} {Models}},
	volume = {24},
	issn = {1061-8600},
	url = {https://www.ncbi.nlm.nih.gov/pmc/articles/PMC4560367/},
	doi = {10.1080/10618600.2014.901914},
	number = {2},
	urldate = {2021-03-09},
	journal = {Journal of Computational and Graphical Statistics},
	author = {Scheipl, Fabian and Staicu, Ana-Maria and Greven, Sonja},
	month = apr,
	year = {2015},
	pmid = {26347592},
	pmcid = {PMC4560367},
	pages = {477--501},
}

@misc{leroux_fast_2023,
	title = {Fast {Generalized} {Functional} {Principal} {Components} {Analysis} {[arXiv:2305.02389 [stat]]}},
	url = {http://arxiv.org/abs/2305.02389},
	doi = {10.48550/arXiv.2305.02389},
	urldate = {2023-07-06},
	publisher = {arXiv},
	author = {Leroux, Andrew and Crainiceanu, M and Wrobel, Julia},
	month = jun,
	year = {2023},
	note = {arXiv:2305.02389 [stat]},
}

@article{greven_longitudinal_2010,
	title = {Longitudinal functional principal component analysis},
	volume = {4},
	issn = {1935-7524},
	url = {https://www.ncbi.nlm.nih.gov/pmc/articles/PMC3131008/},
	doi = {10.1214/10-EJS575},
	urldate = {2020-09-24},
	journal = {Electronic Journal of Statistics},
	author = {Greven, Sonja and Crainiceanu, Ciprian M and Caffo, Brian and Reich, Daniel},
	year = {2010},
	pmid = {21743825},
	pmcid = {PMC3131008},
	pages = {1022--1054},
}

@phdthesis{dillon_investigation_2022,
	type = {Doctoral {Thesis}},
	title = {An investigation of the factors associated with running-related injuries among recreational runners},
	url = {https://doras.dcu.ie/27694/},
	language = {en},
	urldate = {2023-01-14},
	school = {Dublin City University. School of Health and Human Performance},
	author = {Dillon, Sarah},
	month = nov,
	year = {2022},
	note = {Publication Title: Sarah, Dillon ORCID: 0000-0002-6659-2606 {\textless}https://orcid.org/0000-0002-6659-2606{\textgreater}  (2022) An investigation of the factors associated with running-related injuries among recreational runners.  PhD thesis, Dublin City University.},
}

@article{chi_models_1989,
	title = {Models for {Longitudinal} {Data} with {Random} {Effects} and {AR}(1) {Errors}},
	volume = {84},
	issn = {0162-1459},
	url = {https://www.jstor.org/stable/2289929},
	doi = {10.2307/2289929},
	number = {406},
	urldate = {2023-07-15},
	journal = {Journal of the American Statistical Association},
	author = {Chi, Eric M. and Reinsel, Gregory C.},
	year = {1989},
	note = {Publisher: [American Statistical Association, Taylor \& Francis, Ltd.]},
	pages = {452--459},
}

@article{orendurff_little_2018,
	title = {A little bit faster: {Lower} extremity joint kinematics and kinetics as recreational runners achieve faster speeds},
	volume = {71},
	issn = {0021-9290},
	shorttitle = {A little bit faster},
	url = {https://www.sciencedirect.com/science/article/pii/S0021929018300964},
	doi = {10.1016/j.jbiomech.2018.02.010},
	language = {en},
	urldate = {2023-07-12},
	journal = {Journal of Biomechanics},
	author = {Orendurff, Michael S. and Kobayashi, Toshiki and Tulchin-Francis, Kirsten and Tullock, Ann Marie Herring and Villarosa, Chris and Chan, Charles and Kraus, Emily and Strike, Siobhan},
	month = apr,
	year = {2018},
	pages = {167--175},
}

@misc{ramsay_fda_2020,
	title = {{fda}: {Functional} {Data} {Analysis}. {R} package version 5.5.1. {https://CRAN.R-project.org/package=fda}},
	copyright = {GPL-2 {\textbar} GPL-3 [expanded from: GPL (≥ 2)]},
	shorttitle = {fda},
	url = {https://CRAN.R-project.org/package=fda},
	urldate = {2020-10-10},
	author = {Ramsay, James O. and Graves, Spencer and Hooker, Giles},
	month = aug,
	year = {2020},
}

@article{shang_grouped_2017,
	title = {Grouped {Functional} {Time} {Series} {Forecasting}: {An} {Application} to {Age}-{Specific} {Mortality} {Rates}},
	volume = {26},
	issn = {1061-8600},
	shorttitle = {Grouped {Functional} {Time} {Series} {Forecasting}},
	url = {https://doi.org/10.1080/10618600.2016.1237877},
	doi = {10.1080/10618600.2016.1237877},
	number = {2},
	urldate = {2023-07-02},
	journal = {Journal of Computational and Graphical Statistics},
	author = {Shang, Han Lin and Hyndman, Rob J.},
	month = apr,
	year = {2017},
	note = {Publisher: Taylor \& Francis
\_eprint: https://doi.org/10.1080/10618600.2016.1237877},
	pages = {330--343},
}

@misc{bates_parsimonious_2018,
	title = {Parsimonious {Mixed} {Models} {[arXiv:1506.04967v2 [stat]]}},
	url = {http://arxiv.org/abs/1506.04967v2},
	doi = {10.48550/arXiv.1506.04967v2},
	urldate = {2023-06-22},
	publisher = {arXiv},
	author = {Bates, Douglas and Kliegl, Reinhold and Vasishth, Shravan and Baayen, Harald},
	month = may,
	year = {2018},
	note = {arXiv:1506.04967 [stat]},
}

@article{crainiceanu_bootstrap-based_2012,
	title = {Bootstrap-based inference on the difference in the means of two correlated functional processes},
	volume = {31},
	issn = {0277-6715},
	url = {https://www.ncbi.nlm.nih.gov/pmc/articles/PMC3966027/},
	doi = {10.1002/sim.5439},
	number = {26},
	urldate = {2022-01-10},
	journal = {Statistics in Medicine},
	author = {Crainiceanu, Ciprian M. and Staicu, Ana-Maria and Ray, Shubankar and Punjabi, Naresh},
	month = nov,
	year = {2012},
	pmid = {22855258},
	pmcid = {PMC3966027},
	pages = {3223--3240},
}

@article{knief_violating_2021,
	title = {Violating the normality assumption may be the lesser of two evils},
	volume = {53},
	issn = {1554-3528},
	url = {https://doi.org/10.3758/s13428-021-01587-5},
	doi = {10.3758/s13428-021-01587-5},
	language = {en},
	number = {6},
	urldate = {2023-06-15},
	journal = {Behavior Research Methods},
	author = {Knief, Ulrich and Forstmeier, Wolfgang},
	month = dec,
	year = {2021},
	pages = {2576--2590},
}

@article{asar_linear_2020,
	title = {Linear {Mixed} {Effects} {Models} for {Non}-{Gaussian} {Continuous} {Repeated} {Measurement} {Data}},
	volume = {69},
	issn = {0035-9254},
	url = {https://doi.org/10.1111/rssc.12405},
	doi = {10.1111/rssc.12405},
	number = {5},
	urldate = {2023-06-12},
	journal = {Journal of the Royal Statistical Society Series C: Applied Statistics},
	author = {Asar, Özgür and Bolin, David and Diggle, Peter J. and Wallin, Jonas},
	month = nov,
	year = {2020},
	pages = {1015--1065},
}

@article{cui_fast_2023,
	title = {Fast {Multilevel} {Functional} {Principal} {Component} {Analysis}},
	volume = {32},
	issn = {1061-8600},
	url = {https://doi.org/10.1080/10618600.2022.2115500},
	doi = {10.1080/10618600.2022.2115500},
	number = {2},
	urldate = {2023-06-05},
	journal = {Journal of Computational and Graphical Statistics},
	author = {Cui, Erjia and Li, Ruonan and Crainiceanu, Ciprian M. and Xiao, Luo},
	month = apr,
	year = {2023},
	note = {Publisher: Taylor \& Francis
\_eprint: https://doi.org/10.1080/10618600.2022.2115500},
	pages = {366--377},
}

@article{reimherr_functional_2014,
	title = {A {Functional} {Data} {Analysis} {Approach} for {Genetic} {Association} {Studies}},
	volume = {8},
	issn = {1932-6157},
	url = {https://www.jstor.org/stable/24521739},
	number = {1},
	urldate = {2023-05-25},
	journal = {The Annals of Applied Statistics},
	author = {Reimherr, Matthew and Nicolae, Dan},
	year = {2014},
	note = {Publisher: Institute of Mathematical Statistics},
	pages = {406--429},
}

@article{jacqmin-gadda_robustness_2007,
	title = {Robustness of the linear mixed model to misspecified error distribution},
	volume = {51},
	issn = {0167-9473},
	url = {https://www.sciencedirect.com/science/article/pii/S016794730600185X},
	doi = {10.1016/j.csda.2006.05.021},
	language = {en},
	number = {10},
	urldate = {2023-05-05},
	journal = {Computational Statistics \& Data Analysis},
	author = {Jacqmin-Gadda, Hélène and Sibillot, Solenne and Proust, Cécile and Molina, Jean-Michel and Thiébaut, Rodolphe},
	month = jun,
	year = {2007},
	pages = {5142--5154},
}

@misc{gunning_analyzing_2023,
	title = {Analysing {Kinematic} {Data} from {Recreational} {Runners} using {Functional} {Data} {Analysis} [{arXiv}:2408.08200 [stat]]},
	url = {http://arxiv.org/abs/2408.08200},
	doi = {10.48550/arXiv.2408.08200},
	urldate = {2023-06-27},
	publisher = {arXiv},
	author = {Gunning, Edward and Golovkine, Steven and Simpkin, Andrew J. and Burke, Aoife and Dillon, Sarah and Gore, Shane and Moran, Kieran A. and O'Connor, Siobhán and Whyte and Bargary, Norma},
	month = aug,
	year = {2024},
	note = {arXiv:2408.08200 [stat]},
}

@article{boland_study_2022,
	title = {A study of longitudinal trends in time-frequency transformations of {EEG} data during a learning experiment},
	volume = {167},
	issn = {0167-9473},
	url = {https://www.sciencedirect.com/science/article/pii/S0167947321002012},
	doi = {10.1016/j.csda.2021.107367},
	language = {en},
	urldate = {2023-04-14},
	journal = {Computational Statistics \& Data Analysis},
	author = {Boland, Joanna and Telesca, Donatello and Sugar, Catherine and Jeste, Shafali and Goldbeck, Cameron and Şentürk, Damla},
	month = mar,
	year = {2022},
	pages = {107367},
}

@article{shamshoian_bayesian_2022,
	title = {Bayesian analysis of longitudinal and multidimensional functional data},
	volume = {23},
	issn = {1468-4357},
	doi = {10.1093/biostatistics/kxaa041},
	language = {eng},
	number = {2},
	journal = {Biostatistics},
	author = {Shamshoian, John and Şentürk, Damla and Jeste, Shafali and Telesca, Donatello},
	month = apr,
	year = {2022},
	pmid = {33017019},
	pmcid = {PMC9007445},
	pages = {558--573},
}

@article{koner_second-generation_2023,
	title = {Second-{Generation} {Functional} {Data}},
	volume = {10},
	url = {https://doi.org/10.1146/annurev-statistics-032921-033726},
	doi = {10.1146/annurev-statistics-032921-033726},
	number = {1},
	urldate = {2023-04-14},
	journal = {Annual Review of Statistics and Its Application},
	author = {Koner, Salil and Staicu, Ana-Maria},
	year = {2023},
	note = {\_eprint: https://doi.org/10.1146/annurev-statistics-032921-033726},
	pages = {547--572},
}

@article{scheffler_hybrid_2020,
	title = {Hybrid principal components analysis for region-referenced longitudinal functional {EEG} data},
	volume = {21},
	issn = {1465-4644},
	url = {https://doi.org/10.1093/biostatistics/kxy034},
	doi = {10.1093/biostatistics/kxy034},
	number = {1},
	urldate = {2023-04-14},
	journal = {Biostatistics},
	author = {Scheffler, Aaron and Telesca, Donatello and Li, Qian and Sugar, Catherine A and Distefano, Charlotte and Jeste, Shafali and Şentürk, Damla},
	month = jan,
	year = {2020},
	pages = {139--157},
}

@article{chen_modelling_2017,
	title = {Modelling function-valued stochastic processes, with applications to fertility dynamics},
	volume = {79},
	issn = {1369-7412},
	url = {https://www.jstor.org/stable/44681767},
	number = {1},
	urldate = {2023-04-14},
	journal = {Journal of the Royal Statistical Society Series B: Statistical Methodology},
	author = {Chen, Kehui and Delicado, Pedro and Müller, Hans-Georg},
	year = {2017},
	note = {Publisher: [Royal Statistical Society, Wiley]},
	pages = {177--196},
}

@article{hasenstab_multi-dimensional_2017,
	title = {A multi-dimensional functional principal components analysis of {EEG} data},
	volume = {73},
	issn = {1541-0420},
	url = {https://onlinelibrary.wiley.com/doi/abs/10.1111/biom.12635},
	doi = {10.1111/biom.12635},
	language = {en},
	number = {3},
	urldate = {2023-04-14},
	journal = {Biometrics},
	author = {Hasenstab, Kyle and Scheffler, Aaron and Telesca, Donatello and Sugar, Catherine A. and Jeste, Shafali and DiStefano, Charlotte and Şentürk, Damla},
	year = {2017},
	note = {\_eprint: https://onlinelibrary.wiley.com/doi/pdf/10.1111/biom.12635},
	pages = {999--1009},
}

@article{fox_measurement_2023,
	title = {Measurement error associated with gait cycle selection in treadmill running at various speeds},
	volume = {11},
	issn = {2167-8359},
	url = {https://peerj.com/articles/14921},
	doi = {10.7717/peerj.14921},
	language = {en},
	urldate = {2023-04-12},
	journal = {PeerJ},
	author = {Fox, Aaron S. and Bonacci, Jason and Warmenhoven, John and Keast, Meghan F.},
	month = mar,
	year = {2023},
	note = {Publisher: PeerJ Inc.},
	pages = {e14921},
}

@article{park_longitudinal_2015,
	title = {Longitudinal functional data analysis},
	volume = {4},
	issn = {2049-1573},
	url = {https://onlinelibrary.wiley.com/doi/abs/10.1002/sta4.89},
	doi = {10.1002/sta4.89},
	language = {en},
	number = {1},
	urldate = {2023-03-16},
	journal = {Stat},
	author = {Park, So Young and Staicu, Ana-Maria},
	year = {2015},
	note = {\_eprint: https://onlinelibrary.wiley.com/doi/pdf/10.1002/sta4.89},
	pages = {212--226},
}

@article{li_fixed-effects_2022,
	title = {Fixed-effects inference and tests of correlation for longitudinal functional data},
	volume = {41},
	issn = {1097-0258},
	url = {https://onlinelibrary.wiley.com/doi/abs/10.1002/sim.9421},
	doi = {10.1002/sim.9421},
	language = {en},
	number = {17},
	urldate = {2023-02-16},
	journal = {Statistics in Medicine},
	author = {Li, Ruonan and Xiao, Luo and Smirnova, Ekaterina and Cui, Erjia and Leroux, Andrew and Crainiceanu, Ciprian M.},
	year = {2022},
	note = {\_eprint: https://onlinelibrary.wiley.com/doi/pdf/10.1002/sim.9421},
	pages = {3349--3364},
}

@misc{r_core_team_r_2022,
	address = {Vienna, Austria},
	title = {R: {A} {Language} and {Environment} for {Statistical} {Computing}},
	url = {https://www.R-project.org/},
	publisher = {R Foundation for Statistical Computing},
	author = {{R Core Team}},
	year = {2022},
}

@article{di_multilevel_2009,
	title = {Multilevel functional principal component analysis},
	volume = {3},
	issn = {1932-6157},
	url = {https://www.ncbi.nlm.nih.gov/pmc/articles/PMC2835171/},
	doi = {10.1214/08-AOAS206SUPP},
	number = {1},
	urldate = {2020-08-23},
	journal = {The Annals of Applied Statistics},
	author = {Di, Chong-Zhi and Crainiceanu, Ciprian M. and Caffo, Brian S. and Punjabi, Naresh M.},
	month = mar,
	year = {2009},
	pmid = {20221415},
	pmcid = {PMC2835171},
	pages = {458--488},
}

@techreport{vicon_plug-gait_2022,
	type = {Reference {Guide}},
	title = {Plug-{In} {Gait} {Reference} {Guide}},
	url = {https://docs.vicon.com/display/Nexus214/PDF+downloads+for+Vicon+Nexus?preview=/155746642/155746855/Vicon%20Nexus%20Reference%20Guide.pdf},
	urldate = {2023-01-26},
	institution = {Vicon},
	author = {Vicon},
	month = mar,
	year = {2022},
}

@article{fukuchi_public_2017,
	title = {A public dataset of running biomechanics and the effects of running speed on lower extremity kinematics and kinetics},
	volume = {5},
	issn = {2167-8359},
	url = {https://peerj.com/articles/3298},
	doi = {10.7717/peerj.3298},
	language = {en},
	urldate = {2023-01-24},
	journal = {PeerJ},
	author = {Fukuchi, Reginaldo K. and Fukuchi, Claudiane A. and Duarte, Marcos},
	month = may,
	year = {2017},
	note = {Publisher: PeerJ Inc.},
	pages = {e3298},
}

@article{zhu_multivariate_2017,
	title = {Multivariate functional response regression, with application to fluorescence spectroscopy in a cervical pre-cancer study},
	volume = {111},
	issn = {0167-9473},
	url = {https://www.sciencedirect.com/science/article/pii/S0167947317300245},
	doi = {10.1016/j.csda.2017.02.004},
	language = {en},
	urldate = {2023-01-20},
	journal = {Computational Statistics \& Data Analysis},
	author = {Zhu, Hongxiao and Morris, Jeffrey S. and Wei, Fengrong and Cox, Dennis D.},
	month = jul,
	year = {2017},
	pages = {88--101},
}

@article{taylor_repeatability_2010,
	title = {Repeatability and reproducibility of {OSSCA}, a functional approach for assessing the kinematics of the lower limb},
	volume = {32},
	issn = {0966-6362},
	url = {https://www.sciencedirect.com/science/article/pii/S0966636210001293},
	doi = {10.1016/j.gaitpost.2010.05.005},
	language = {en},
	number = {2},
	urldate = {2023-01-17},
	journal = {Gait \& Posture},
	author = {Taylor, W. R. and Kornaropoulos, E. I. and Duda, G. N. and Kratzenstein, S. and Ehrig, R. M. and Arampatzis, A. and Heller, M. O.},
	month = jun,
	year = {2010},
	pages = {231--236},
}

@article{goldsmith_generalized_2015,
	title = {Generalized multilevel function-on-scalar regression and principal component analysis},
	volume = {71},
	issn = {1541-0420},
	url = {https://onlinelibrary.wiley.com/doi/abs/10.1111/biom.12278},
	doi = {10.1111/biom.12278},
	language = {en},
	number = {2},
	urldate = {2023-01-13},
	journal = {Biometrics},
	author = {Goldsmith, Jeff and Zipunnikov, Vadim and Schrack, Jennifer},
	year = {2015},
	note = {\_eprint: https://onlinelibrary.wiley.com/doi/pdf/10.1111/biom.12278},
	pages = {344--353},
}

@article{zhao_functional_2004,
	title = {The {Functional} {Data} {Analysis} {View} of {Longitudinal} {Data}},
	volume = {14},
	issn = {1017-0405},
	url = {https://www.jstor.org/stable/24307416},
	number = {3},
	urldate = {2023-01-13},
	journal = {Statistica Sinica},
	author = {Zhao, Xin and Marron, J. S. and Wells, Martin T.},
	year = {2004},
	note = {Publisher: Institute of Statistical Science, Academia Sinica},
	pages = {789--808},
}

@article{muller_functional_2005,
	title = {Functional {Modelling} and {Classification} of {Longitudinal} {Data}},
	volume = {32},
	issn = {1467-9469},
	url = {https://onlinelibrary.wiley.com/doi/abs/10.1111/j.1467-9469.2005.00429.x},
	doi = {10.1111/j.1467-9469.2005.00429.x},
	language = {en},
	number = {2},
	urldate = {2023-01-13},
	journal = {Scandinavian Journal of Statistics},
	author = {Müller, Hans-Georg},
	year = {2005},
	note = {\_eprint: https://onlinelibrary.wiley.com/doi/pdf/10.1111/j.1467-9469.2005.00429.x},
	pages = {223--240},
}

@article{rice_functional_2004,
	title = {Functional and {Longitudinal} {Data} {Analysis}: {Perspectives} on {Smoothing}},
	volume = {14},
	issn = {1017-0405},
	shorttitle = {Functional and {Longitudinal} {Data} {Analysis}},
	url = {https://www.jstor.org/stable/24307409},
	number = {3},
	urldate = {2023-01-13},
	journal = {Statistica Sinica},
	author = {Rice, John A.},
	year = {2004},
	note = {Publisher: Institute of Statistical Science, Academia Sinica},
	pages = {631--647},
}

@article{park_simple_2018,
	title = {Simple fixed-effects inference for complex functional models},
	volume = {19},
	issn = {1465-4644},
	url = {https://doi.org/10.1093/biostatistics/kxx026},
	doi = {10.1093/biostatistics/kxx026},
	number = {2},
	urldate = {2022-10-22},
	journal = {Biostatistics},
	author = {Park, So Young and Staicu, Ana-Maria and Xiao, Luo and Crainiceanu, Ciprian M},
	month = apr,
	year = {2018},
	pages = {137--152},
}

@article{zhang_functional_2016,
	title = {Functional {CAR} models for large spatially correlated functional datasets},
	volume = {111},
	issn = {0162-1459},
	doi = {10.1080/01621459.2015.1042581},
	language = {eng},
	number = {514},
	journal = {Journal of the American Statistical Association},
	author = {Zhang, Lin and Baladandayuthapani, Veerabhadran and Zhu, Hongxiao and Baggerly, Keith A. and Majewski, Tadeusz and Czerniak, Bogdan A. and Morris, Jeffrey S.},
	year = {2016},
	pmid = {28018013},
	pmcid = {PMC5176110},
	pages = {772--786},
}

@article{glazier_beyond_2021,
	title = {Beyond animated skeletons: {How} can biomechanical feedback be used to enhance sports performance?},
	volume = {129},
	issn = {0021-9290},
	shorttitle = {Beyond animated skeletons},
	url = {https://www.sciencedirect.com/science/article/pii/S0021929021004553},
	doi = {10.1016/j.jbiomech.2021.110686},
	language = {en},
	urldate = {2021-10-13},
	journal = {Journal of Biomechanics},
	author = {Glazier, Paul S.},
	month = dec,
	year = {2021},
	pages = {110686},
}

@article{cui_fast_2022,
	title = {Fast {Univariate} {Inference} for {Longitudinal} {Functional} {Models}},
	volume = {31},
	issn = {1061-8600},
	url = {https://doi.org/10.1080/10618600.2021.1950006},
	doi = {10.1080/10618600.2021.1950006},
	number = {1},
	urldate = {2022-05-31},
	journal = {Journal of Computational and Graphical Statistics},
	author = {Cui, Erjia and Leroux, Andrew and Smirnova, Ekaterina and Crainiceanu, Ciprian M.},
	month = jan,
	year = {2022},
	note = {Publisher: Taylor \& Francis
\_eprint: https://doi.org/10.1080/10618600.2021.1950006},
	pages = {219--230},
}

@article{hadjipantelis_unifying_2015,
	title = {Unifying {Amplitude} and {Phase} {Analysis}: {A} {Compositional} {Data} {Approach} to {Functional} {Multivariate} {Mixed}-{Effects} {Modeling} of {Mandarin} {Chinese}},
	volume = {110},
	issn = {0162-1459},
	shorttitle = {Unifying {Amplitude} and {Phase} {Analysis}},
	url = {https://doi.org/10.1080/01621459.2015.1006729},
	doi = {10.1080/01621459.2015.1006729},
	number = {510},
	urldate = {2022-05-08},
	journal = {Journal of the American Statistical Association},
	author = {Hadjipantelis, P. Z. and Aston, J. A. D. and Müller, H. G. and Evans, J. P.},
	month = apr,
	year = {2015},
	pmid = {26692591},
	note = {Publisher: Taylor \& Francis
\_eprint: https://doi.org/10.1080/01621459.2015.1006729},
	pages = {545--559},
}

@article{lee_bayesian_2019,
	title = {Bayesian {Semiparametric} {Functional} {Mixed} {Models} for {Serially} {Correlated} {Functional} {Data}, {With} {Application} to {Glaucoma} {Data}},
	volume = {114},
	issn = {0162-1459},
	url = {https://doi.org/10.1080/01621459.2018.1476242},
	doi = {10.1080/01621459.2018.1476242},
	number = {526},
	urldate = {2022-04-28},
	journal = {Journal of the American Statistical Association},
	author = {Lee, Wonyul and Miranda, Michelle F. and Rausch, Philip and Baladandayuthapani, Veerabhadran and Fazio, Massimo and Downs, J. Crawford and Morris, Jeffrey S.},
	month = apr,
	year = {2019},
	pmid = {31235987},
	note = {Publisher: Taylor \& Francis
\_eprint: https://doi.org/10.1080/01621459.2018.1476242},
	pages = {495--513},
}

@article{aston_linguistic_2010,
	title = {Linguistic {Pitch} {Analysis} using {Functional} {Principal} {Component} {Mixed} {Effect} {Models}},
	volume = {59},
	issn = {0035-9254},
	url = {https://www.jstor.org/stable/40541687},
	number = {2},
	urldate = {2022-03-01},
	journal = {Journal of the Royal Statistical Society Series C: Applied Statistics},
	author = {Aston, John A. D. and Chiou, Jeng-Min and Evans, Jonathan P.},
	year = {2010},
	note = {Publisher: [Wiley, Royal Statistical Society]},
	pages = {297--317},
}

@article{morris_automated_2011,
	title = {Automated analysis of quantitative image data using isomorphic functional mixed models, with application to proteomics data},
	volume = {5},
	issn = {1932-6157, 1941-7330},
	url = {https://projecteuclid.org/journals/annals-of-applied-statistics/volume-5/issue-2A/Automated-analysis-of-quantitative-image-data-using-isomorphic-functional-mixed/10.1214/10-AOAS407.full},
	doi = {10.1214/10-AOAS407},
	number = {2A},
	urldate = {2022-03-01},
	journal = {The Annals of Applied Statistics},
	author = {Morris, Jeffrey S. and Baladandayuthapani, Veerabhadran and Herrick, Richard C. and Sanna, Pietro and Gutstein, Howard},
	month = jun,
	year = {2011},
	note = {Publisher: Institute of Mathematical Statistics},
	pages = {894--923},
}

@misc{volkmann_multifamm_2021,
	title = {{multifamm}: {Multivariate} {Functional} {Additive} {Mixed} {Models}. {R} package version 0.1.1. {https://CRAN.R-project.org/package=multi} {famm}},
	copyright = {GPL-2 {\textbar} GPL-3 [expanded from: GPL (≥ 2)]},
	shorttitle = {multifamm},
	url = {https://CRAN.R-project.org/package=multifamm},
	urldate = {2022-02-08},
	author = {Volkmann, Alexander},
	month = sep,
	year = {2021},
}

@article{bauer_introduction_2018,
	title = {An introduction to semiparametric function-on-scalar regression},
	volume = {18},
	issn = {1471-082X},
	url = {https://doi.org/10.1177/1471082X17748034},
	doi = {10.1177/1471082X17748034},
	language = {en},
	number = {3-4},
	urldate = {2021-11-25},
	journal = {Statistical Modelling},
	author = {Bauer, Alexander and Scheipl, Fabian and Küchenhoff, Helmut and Gabriel, Alice-Agnes},
	month = jun,
	year = {2018},
	note = {Publisher: SAGE Publications India},
	pages = {346--364},
}

@article{morris_comparison_2017,
	title = {Comparison and contrast of two general functional regression modelling frameworks},
	volume = {17},
	issn = {1471-082X},
	url = {https://doi.org/10.1177/1471082X16681875},
	doi = {10.1177/1471082X16681875},
	language = {en},
	number = {1-2},
	urldate = {2021-11-25},
	journal = {Statistical Modelling},
	author = {Morris, Jeffrey S.},
	month = feb,
	year = {2017},
	note = {Publisher: SAGE Publications India},
	pages = {59--85},
}

@article{morris_wavelet-based_2006,
	title = {Wavelet-based functional mixed models},
	volume = {68},
	issn = {1369-7412},
	doi = {10.1111/j.1467-9868.2006.00539.x},
	language = {eng},
	number = {2},
	journal = {Journal of the Royal Statistical Society Series B: Statistical Methodology},
	author = {Morris, Jeffrey S. and Carroll, Raymond J.},
	month = apr,
	year = {2006},
	pmid = {19759841},
	pmcid = {PMC2744105},
	pages = {179--199},
}

@article{chen_modeling_2012,
	title = {Modeling {Repeated} {Functional} {Observations}},
	volume = {107},
	issn = {0162-1459},
	url = {https://www.jstor.org/stable/23427358},
	number = {500},
	urldate = {2021-05-14},
	journal = {Journal of the American Statistical Association},
	author = {Chen, Kehui and Müller, Hans-Georg},
	year = {2012},
	note = {Publisher: [American Statistical Association, Taylor \& Francis, Ltd.]},
	pages = {1599--1609},
}

@article{laird_random-effects_1982,
	title = {Random-{Effects} {Models} for {Longitudinal} {Data}},
	volume = {38},
	issn = {0006-341X},
	url = {https://www.jstor.org/stable/2529876},
	doi = {10.2307/2529876},
	number = {4},
	urldate = {2021-05-14},
	journal = {Biometrics},
	author = {Laird, Nan M. and Ware, James H.},
	year = {1982},
	note = {Publisher: [Wiley, International Biometric Society]},
	pages = {963--974},
}

@article{hyndman_rainbow_2010,
	title = {Rainbow {Plots}, {Bagplots}, and {Boxplots} for {Functional} {Data}},
	volume = {19},
	issn = {1061-8600},
	url = {https://www.jstor.org/stable/25651298},
	number = {1},
	urldate = {2021-02-26},
	journal = {Journal of Computational and Graphical Statistics},
	author = {Hyndman, Rob J. and Shang, Han Lin},
	year = {2010},
	note = {Publisher: [American Statistical Association, Taylor \& Francis, Ltd., Institute of Mathematical Statistics, Interface Foundation of America]},
	pages = {29--45},
}

@article{happ_multivariate_2018,
	title = {Multivariate {Functional} {Principal} {Component} {Analysis} for {Data} {Observed} on {Different} ({Dimensional}) {Domains}},
	volume = {113},
	issn = {0162-1459},
	url = {https://doi.org/10.1080/01621459.2016.1273115},
	doi = {10.1080/01621459.2016.1273115},
	number = {522},
	urldate = {2021-02-10},
	journal = {Journal of the American Statistical Association},
	author = {Happ, Clara and Greven, Sonja},
	month = apr,
	year = {2018},
	note = {Publisher: Taylor \& Francis
\_eprint: https://doi.org/10.1080/01621459.2016.1273115},
	pages = {649--659},
}

@article{bates_fitting_2015,
	title = {Fitting {Linear} {Mixed}-{Effects} {Models} {Using} lme4},
	volume = {67},
	copyright = {Copyright (c) 2015 Douglas Bates, Martin Mächler, Ben Bolker, Steve Walker},
	issn = {1548-7660},
	url = {https://www.jstatsoft.org/index.php/jss/article/view/v067i01},
	doi = {10.18637/jss.v067.i01},
	language = {en},
	number = {1},
	urldate = {2020-12-21},
	journal = {Journal of Statistical Software},
	author = {Bates, Douglas and Mächler, Martin and Bolker, Ben and Walker, Steve},
	month = oct,
	year = {2015},
	note = {Number: 1},
	pages = {1--48},
}

@article{yao_functional_2005,
	title = {Functional {Data} {Analysis} for {Sparse} {Longitudinal} {Data}},
	volume = {100},
	issn = {0162-1459},
	url = {https://www.jstor.org/stable/27590579},
	number = {470},
	urldate = {2020-10-12},
	journal = {Journal of the American Statistical Association},
	author = {Yao, Fang and Müller, Hans-Georg and Wang, Jane-Ling},
	year = {2005},
	note = {Publisher: [American Statistical Association, Taylor \& Francis, Ltd.]},
	pages = {577--590},
}

@article{jacques_model-based_2014,
	title = {Model-based clustering for multivariate functional data},
	volume = {71},
	issn = {0167-9473},
	url = {http://www.sciencedirect.com/science/article/pii/S0167947312004380},
	doi = {10.1016/j.csda.2012.12.004},
	language = {en},
	urldate = {2020-09-22},
	journal = {Computational Statistics \& Data Analysis},
	author = {Jacques, Julien and Preda, Cristian},
	month = mar,
	year = {2014},
	pages = {92--106},
}

@article{park_functional_2017,
	title = {Functional vs. {Traditional} {Analysis} in {Biomechanical} {Gait} {Data}: {An} {Alternative} {Statistical} {Approach}},
	volume = {60},
	issn = {1640-5544},
	shorttitle = {Functional vs. {Traditional} {Analysis} in {Biomechanical} {Gait} {Data}},
	doi = {10.1515/hukin-2017-0114},
	language = {eng},
	journal = {Journal of Human Kinetics},
	author = {Park, Jihong and Seeley, Matthew K. and Francom, Devin and Reese, C. Shane and Hopkins, J. Ty},
	month = dec,
	year = {2017},
	pmid = {29339984},
	pmcid = {PMC5765784},
	pages = {39--49},
}

\clearpage
\appendix
\section{Implementation Details}
All analyses were performed in \proglang{R} version 4.1.2 \parencite{r_core_team_r_2022}. The \pkg{fda} \parencite{ramsay_fda_2020} package was used for the basis expansion and mv-FPCA steps. The \pkg{lme4} package \parencite{bates_fitting_2015} was used to fit the univariate scalar mixed effects models. 
The data analysis was performed on a 2019 MacBook Pro with a 2.4 GHz Quad-Core Intel Core i5 processor and 8 GB of memory. 
The simulation was performed on the Irish Centre for High-End Computing (ICHEC) cluster, with 1 core per simulation replicate.
We have prepared a GitHub repository containing custom functions to implement our methods and scripts to reproduce the results of the data analysis and simulations contained in the manuscript, which is available at \url{https://github.com/FAST-ULxNUIG/RISC1-longitudinal-manuscript-code}.

\clearpage
\section{Covariance Functions}

\subsection{Basis Representation of the Covariance Functions}\label{sec:basis-representation-of-covariance}

\subsubsection{Subject-Level Covariance}
Given $\mathbf{u}_i (t, T)  = \sum_{k=1}^K \sum_{d=1}^D u^*_{i, k, d} \ \xi_d(T) \boldpsi_k(t)$, we have
$$
\mathbf{u}_i (t, T)  \mathbf{u}_i (t', T')^\top = \sum_{k=1}^K \sum_{k'=1}^K \sum_{d=1}^D \sum_{d'=1}^D u^*_{i, k, d} \  u^*_{i, k', d'} \xi_d(T)  \ \xi_{d'}(T')  \boldpsi_k(t) \boldpsi_{k'}(t')^\top.
$$
Then 
\begin{align}
    \mathbf{Q}(t, t', T, T') &=  \Expec\left[\mathbf{u}_i(t, T)  \mathbf{u}_i(t', T')^\top\right] \\
    &= \sum_{k=1}^K \sum_{k'=1}^K \sum_{d=1}^D \sum_{d'=1}^D \Expec[u^*_{i, k, d} \  u^*_{i, k', d'}] \ \xi_d(T)  \ \xi_{d'}(T')  \boldpsi_k(t) \boldpsi_{k'}(t')^\top \\
    &= \sum_{k=1}^K \sum_{k'=1}^K \sum_{d=1}^D \sum_{d'=1}^D \Cov\bigl(u^*_{i, k, d},  u^*_{i, k', d'} \bigr) \ \xi_d(T)  \ \xi_{d'}(T')  \boldpsi_k(t) \boldpsi_{k'}(t')^\top \\
    &= \sum_{k=1}^K \sum_{d=1}^D \sum_{d'=1}^D \Cov\bigl(u^*_{i, k, d},   u^*_{i, k, d'} \bigr) \ \xi_d(T)  \ \xi_{d'}(T')  \boldpsi_k(t) \boldpsi_{k}(t')^\top, \\
\end{align}
because $\Cov\bigl(u^*_{i, k, d},  u^*_{i, k', d'} \bigr) = 0$ for $k \neq k'$ due to the assumption of independence across the basis coefficients. Letting $\boldsymbol{\xi} (T) = (\xi_1(T), \dots, \xi_D (T))^\top$, this can be re-written as
\begin{equation}\label{eq:sandwich-basis}
    \sum_{k=1}^K \boldpsi_k(t) \boldpsi_{k}(t')^\top \boldsymbol{\xi} (T)^\top  \mathbf{Q}_k^* \ \boldsymbol{\xi} (T') =  \boldsymbol{\Psi} (t)^\top \mathbf{Q}^*(T, T') \boldsymbol{\Psi} (t'),
\end{equation}
where $\mathbf{Q}^*(T, T') = \diag\{\boldsymbol{\xi} (T)^\top \mathbf{Q}_1^* \boldsymbol{\xi} (T'), \dots, \boldsymbol{\xi} (T)^\top \mathbf{Q}_K^* \boldsymbol{\xi} (T') \}$. Using the Kronecker product $\otimes$, we re-write
$$
\mathbf{Q}^*(T, T') = (\mathbf{I}_K \otimes \boldsymbol{\xi}(T))^\top \mathbf{Q}^* (\mathbf{I}_K \otimes \boldsymbol{\xi}(T')), 
$$
where $\mathbf{Q}^*$ is the block-diagonal matrix containing the matrices $\mathbf{Q}^*_1, \dots, \mathbf{Q}^*_K$ along its diagonal. Subbing this quantity back into \eqref{eq:sandwich-basis} gives the expression 
\begin{align*}
     \mathbf{Q}(t, t', T, T') &= \boldsymbol{\Psi} (t)^\top (\mathbf{I}_K \otimes \boldsymbol{\xi}(T))^\top \mathbf{Q}^* (\mathbf{I}_K \otimes \boldsymbol{\xi}(T'))  \boldsymbol{\Psi} (t') \\
     &= 
    \bigl((\mathbf{I}_K \otimes \boldsymbol{\xi}(T)) \boldsymbol{\Psi} (t) \bigr)^\top \mathbf{Q}^* \bigl( (\mathbf{I}_K \otimes \boldsymbol{\xi}(T'))  \boldsymbol{\Psi} (t') \bigr).
\end{align*}

\subsubsection{Subject-and-Side-Level Covariance}

Likewise, for the subject-and-side-level random effects, we have $\mathbf{v}_{ij}(t, T) \allowbreak = \allowbreak \sum_{k=1}^K \sum_{d=1}^D v_{ij, k, d}^*  \ \xi_d(T) \boldpsi_k(t)$, so that
$$
\mathbf{v}_{ij}(t, T) \mathbf{v}_{ij}(t', T')^\top
=
\sum_{k=1}^K \sum_{k'=1}^K \sum_{d=1}^D \sum_{d'=1}^D v^*_{ij, k, d} \  v^*_{ij, k', d'} \xi_d(T)  \ \xi_{d'}(T')  \boldpsi_k(t) \boldpsi_{k'}(t')^\top.
$$
Using the same argument as above, we have
\begin{align}
    \mathbf{R}(t, t', T, T') &=  \Expec \left[\mathbf{v}_{ij}(t, T) \mathbf{v}_{ij}(t', T')^\top \right] \\
    &= \sum_{k=1}^K \sum_{d=1}^D \sum_{d'=1}^D \Cov\bigl(v^*_{ij, k, d},   v^*_{ij, k, d'} \bigr) \ \xi_d(T)  \ \xi_{d'}(T')  \boldpsi_k(t) \boldpsi_{k}(t')^\top. \\
\end{align}
Re-writing into vector and matrix form in a similar manner to above, we obtain
\begin{align*}
     \mathbf{R}(t, t', T, T') &= \boldsymbol{\Psi} (t)^\top (\mathbf{I}_K \otimes \boldsymbol{\xi}(T))^\top \mathbf{R}^* (\mathbf{I}_K \otimes \boldsymbol{\xi}(T'))  \boldsymbol{\Psi} (t') \\
     &= 
    \bigl((\mathbf{I}_K \otimes \boldsymbol{\xi}(T)) \boldsymbol{\Psi} (t) \bigr)^\top \mathbf{R}^* \bigl( (\mathbf{I}_K \otimes \boldsymbol{\xi}(T'))  \boldsymbol{\Psi} (t') \bigr),
\end{align*}
where, again, $\mathbf{R}^*$ is block diagonal matrix and contains $\mathbf{R}^*_1, \dots, \mathbf{R}^*_K$ along its diagonal.

\subsubsection{Within-Function Covariance}

Finally, the multivariate functional random error term can be written as $\boldsymbol{\varepsilon}_{ijl} (t) = \sum_{k=1}^K \varepsilon_{ijl, k}^* \boldpsi_k (t)$ and then
$$
\boldsymbol{\varepsilon}_{ijl} (t) \ \boldsymbol{\varepsilon}_{ijl} (t') ^\top = \sum_{k=1}^K  \sum_{k'=1}^K \varepsilon_{ijl, k}^* \varepsilon_{ijl, k'}^* \boldpsi_k (t) \boldpsi_{k'} (t')^\top.
$$
Then, the within-function covariance can be written as
\begin{align*}
    \mathbf{S}(t, t') &= \Expec \left[\boldsymbol{\varepsilon}_{ijl} (t) \ \boldsymbol{\varepsilon}_{ijl} (t') ^\top \right] \\
    &=
    \sum_{k=1}^K  \sum_{k'=1}^K \Expec[\varepsilon_{ijl, k}^* \varepsilon_{ijl, k'}^*] \boldpsi_k (t) \boldpsi_{k'} (t')^\top \\
    &=
    \sum_{k=1}^K  \sum_{k'=1}^K \Cov \bigl( \varepsilon_{ijl, k}^*, \varepsilon_{ijl, k'}^* \bigr) \boldpsi_k (t) \boldpsi_{k'} (t')^\top \\
    &=
    \sum_{k=1}^K  \underbrace{\Cov \bigl( \varepsilon_{ijl, k}^*,\varepsilon_{ijl, k}^* \bigr)}_{=s_k} \boldpsi_k (t) \boldpsi_{k} (t')^\top \\
    &= 
    \sum_{k=1}^K  s_k \boldpsi_k (t) \boldpsi_{k} (t')^\top \\
    &=
    \boldsymbol{\Psi} (t) ^\top \mathbf{S}^*  \boldsymbol{\Psi} (t'),
\end{align*}
where $\mathbf{S}^* = \diag\{s_1, \dots, s_K\}$.

\subsection{Implied Covariance Between Observations}

With a slight abuse of notation, we let $\mathbf{y}_{ijl} (t)$ denote an observation that has been centered around the fixed effects, i.e., $\boldbeta_0(t, T_{ijl}) + \sum_{a=1}^A x_{ija} \boldbeta_{a} (t)$ has been subtracted. Then, we have
\begin{align*}
    \Cov \bigl( \mathbf{y}_{ijl} (t), \mathbf{y}_{i'j'l'} (t') \bigr) &= \Expec \bigl[ \mathbf{y}_{ijl} (t) \mathbf{y}_{i'j'l'} (t')^\top \bigr] \\
    &= 
    \underbrace{\Expec \bigl[ \mathbf{u}_i (t, T_{ijl}) \mathbf{u}_{i'} (t', T_{i'j'l'})^\top \bigr]}_{= \mathbf{Q}(t, t', T_{ijl}, T_{ij'l'}) \delta_{ii'}} +  \underbrace{\Expec \bigl[ \mathbf{v}_{ij} (t, T_{ijl}) \mathbf{v}_{i'j'} (t', T_{i'j'l'})^\top \bigr]}_{= \mathbf{R}(t, t', T_{ijl}, T_{ijl'}) \delta_{ii'} \delta_{jj'}} \\
    &\quad + \underbrace{\Expec \bigl[ \boldsymbol{\varepsilon}_{ijl} (t) \boldsymbol{\varepsilon}_{i'j'l'} (t')^\top \bigr]}_{= \mathbf{S}(t, t') \delta_{ii'} \delta_{jj'} \delta_{ll'}}, 
\end{align*}
where the simplification arises because the processes $\mathbf{u}_i(t, T) $, $\mathbf{v}_{ij}(t, T)$ and $\boldsymbol{\varepsilon}_{ijl}(t)$ are mutually uncorrelated. Therefore, we have
$$
\Cov \bigl( \mathbf{y}_{ijl} (t), \mathbf{y}_{i'j'l'} (t') \bigr)
=
\begin{cases}
  \mathbf{Q}(t, t', T_{ijl}, T_{ij'l'}) & \text{if $i = i'$, $j \neq j'$ and $l \neq l'$}, \\
  \mathbf{Q}(t, t', T_{ijl}, T_{ijl'}) + \mathbf{R}(t, t', T_{ijl}, T_{ijl'}) & \text{if $i = i'$, $j = j'$ and $l \neq l'$}, \\
  \mathbf{Q}(t, t', T_{ijl}, T_{ijl}) + \mathbf{R}(t, t', T_{ijl}, T_{ijl}) + \mathbf{S}(t, t') & \text{if $i = i'$, $j = j'$ and $l = l'$}, \\
  \mathbf{0} & \text{otherwise}. \\
\end{cases}
$$
Thus, the model is explicitly accounting for multivariate functional dependence along the longitudinal timescale. Take, for example, the multivariate functional observations from two strides from the same subject and side $\mathbf{y}_{ijl} (t)$ and $\mathbf{y}_{ijl'} (t)$, $l \neq l'$. The dependence between the observations depends on the times $T_{ijl}$ and  $T_{ijl'}$ at which the strides occur. The form of this dependence is induced by the basis functions in both the functional and longitudinal directions, as illustrated in Appendix \ref{sec:basis-representation-of-covariance}.

\clearpage
 \section{Additional Simulation Details}

\subsection{Simulation Setup}\label{sec:appendix-simulation-settings}

Figure \ref{fig:psi-simulation} displays the first 10 mv-FPCs used to generate the multivariate functional data in the simulation. The base \proglang{R} function \texttt{poly()} was used to construct the polynomial basis functions in the longitudinal direction that are orthogonalised on an equally-spaced grid of length $101$ on $[0, 1]$. They are defined, based on a recursive formula, as
\begin{align*}
    \xi_{1} (T) &= \frac{1}{\sqrt{101}} \\
    \xi_{2} (T) &= \frac{T - 0.5}{\sqrt{8.585}} \\
    \xi_{3} (T) &= \frac{(T - 0.5)^2 - (8.585/101)}{\sqrt{0.5836083}}.
\end{align*}
However, we used $\xi_{1} (T) = 1$ to align with standard convention for including an intercept in the \proglang{R} software. 
Table \ref{tab:fixed-parameters} contains the basis coefficients used to generate the fixed effects and Table \ref{tab:random-parameters} contains the parameters used to define the distributions of the basis coefficients of the random effects (off-diagonal elements of the covariance matrices were set at $0$). 
All empirical parameters used in the simulation were rounded to the nearest whole number for ease of presentation. 
Figure \ref{fig:simulated-trajectories} displays simulated trajectories of the first basis coefficient (i.e., the first simulated mv-FPC score, labelled mv-FPC1) for six subjects.
As a variance explained cutoff of $99.5\%$ was used, smooth Gaussian noise proportional to the remaining $0.5\%$ was added independently to each dimension of each generated observation \parencite{aston_linguistic_2010}. 
The smooth noise was generated by drawing realisations of a mean-zero Gaussian process with covariance function $C(t, t') = \sigma^2 f( l  \lvert t- t' \rvert)$, where $f(x)$ is the standard Gaussian density.
The parameter values $l=0.25$ and $\sigma = 0.9$ were used.
Figure \ref{fig:simulated-mvmlfd} displays simulated multivariate functional observations for three subjects.

\begin{figure}[h]
    \centering
    \includegraphics[page = 5, width = 0.9\textwidth]{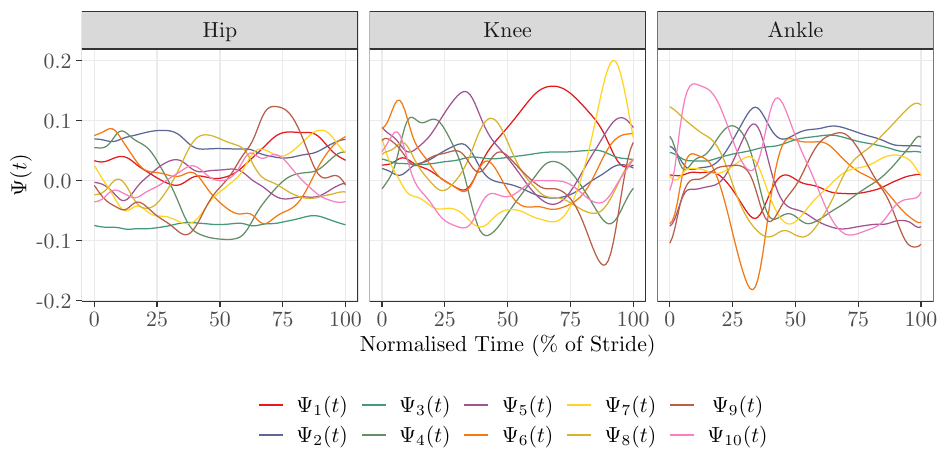}
    \caption{The first 10 empirical mv-FPCs used to generate the data in the simulation.}
    \label{fig:psi-simulation}
\end{figure}

% latex table generated in R 4.1.2 by xtable 1.8-4 package
% Fri Sep  8 16:45:45 2023
\begin{table}[ht]
\centering
\begin{tabular}{lrrrrrrrrrr}
  \toprule
 & {\bfseries 1} & {\bfseries 2} & {\bfseries 3} & {\bfseries 4} & {\bfseries 5} & {\bfseries 6} & {\bfseries 7} & {\bfseries 8} & {\bfseries 9} & {\bfseries 10} \\ 
  \midrule
$\beta_{0, 1, k}^*$ & -3 & 15 & -0 & -3 & 1 & 0 & -4 & -1 & -4 & 1 \\ 
  $\beta_{0, 2, k}^*$ & -4 & -8 & -4 & 4 & 4 & -0 & 5 & -2 & -3 & 4 \\ 
  $\beta_{0, 3, k}^*$ & 6 & 1 & 2 & -1 & -0 & -2 & -1 & 2 & 0 & -1 \\ 
  $\beta_{1, k}^* $ & 11 & -40 & -1 & 9 & -4 & -2 & 10 & 2 & 10 & -2 \\ 
  $\beta_{2, k}^* $ & 28 & -9 & 0 & 5 & -2 & 2 & 1 & 1 & 3 & -2 \\ 
   \bottomrule
\end{tabular}
\caption{Values of the fixed effects basis coefficients used in the simulation.} 
\label{tab:fixed-parameters}
\end{table}

% latex table generated in R 4.1.2 by xtable 1.8-4 package
% Fri Sep  8 17:01:16 2023
\begin{table}[ht]
\centering
\begin{tabular}{lrrrrrrrrrr}
  \toprule
 & {\bfseries 1} & {\bfseries 2} & {\bfseries 3} & {\bfseries 4} & {\bfseries 5} & {\bfseries 6} & {\bfseries 7} & {\bfseries 8} & {\bfseries 9} & {\bfseries 10} \\ 
  \midrule
$q_{11, k}$ & 2590 & 2063 & 1734 & 520 & 417 & 286 & 228 & 194 & 132 & 96 \\ 
  $q_{22, k}$ & 2802 & 890 & 311 & 435 & 279 & 151 & 129 & 120 & 69 & 65 \\ 
  $q_{33, k}$ & 1244 & 382 & 181 & 174 & 116 & 102 & 65 & 53 & 35 & 37 \\ 
  $r_{11, k}$ & 775 & 776 & 368 & 125 & 167 & 79 & 85 & 89 & 49 & 54 \\ 
  $r_{22, k}$ & 317 & 216 & 170 & 80 & 93 & 64 & 21 & 65 & 35 & 48 \\ 
  $r_{33, k}$ & 111 & 82 & 38 & 17 & 53 & 37 & 6 & 29 & 41 & 23 \\ 
  $s_k$ & 135 & 76 & 47 & 58 & 40 & 39 & 51 & 27 & 27 & 18 \\ 
   \bottomrule
\end{tabular}
\caption{Values of the random effects and random error parameters used in the simulation.} 
\label{tab:random-parameters}
\end{table}

\begin{figure}[h]
    \centering
    \includegraphics[page = 6, width = 0.9\textwidth]{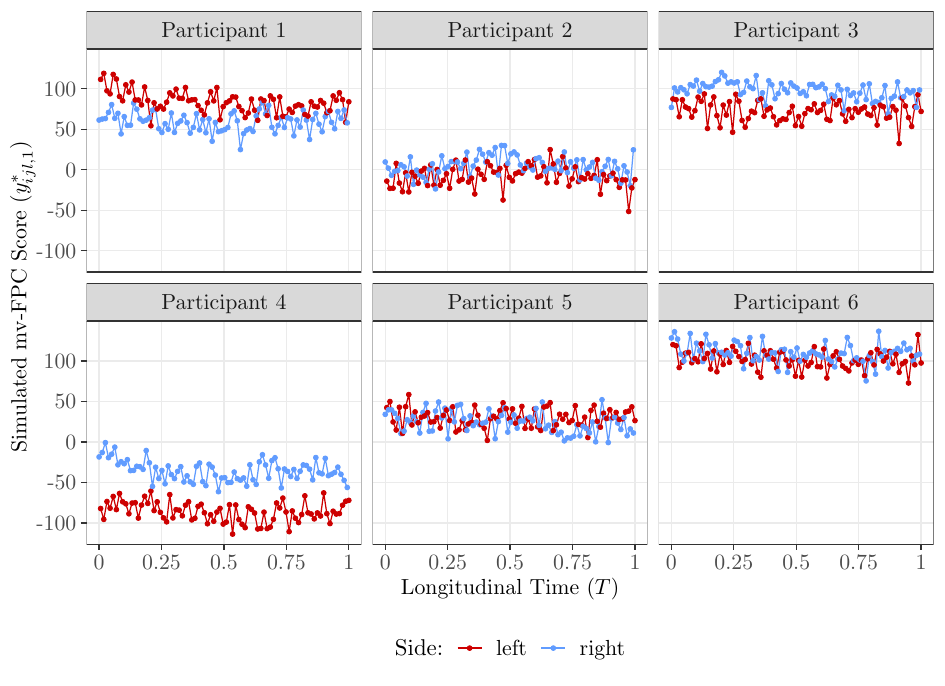}
    \caption{Simulated trajectories of the first basis coefficient for six subjects at $n_{ij} = 80$ equally-spaced points on $[0,1]$.}
    \label{fig:simulated-trajectories}
\end{figure}

\begin{figure}[h]
    \centering
    \includegraphics[page = 3, width = 0.9\textwidth]{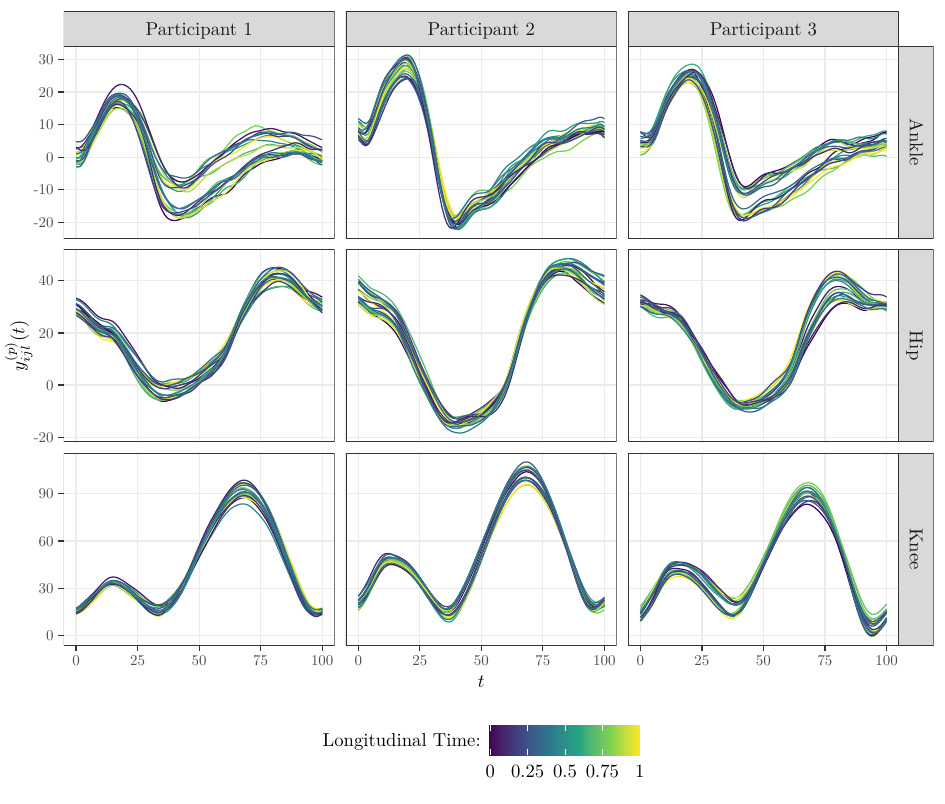}
    \caption{Simulated multivariate functional data for three subjects. To avoid over-plotting and aid visualisation, the data are simulated at $n_{ij} = 10$ (rather than 80) equally-spaced points on $[0, 1]$.}
    \label{fig:simulated-mvmlfd}
\end{figure}

\subsection{Eigenfunction Estimation}\label{app:eigenfunction-recovery}

This section contains a brief description of the eigenfunctions recovered in the simulation. 
Our aim is to highlight, via a short simulation, that the eigenfunctions recovered from the pooled mv-FPCA of the simulated data are linear combinations of the basis functions used to generate the data.
% Our aim is to highlight that, when covariates are used to generate the basis coefficients in the data-generating model \eqref{eq:data-generating-model}, 
This occurs because covariate effects produce small but non-zero marginal correlations among the simulated basis coefficients.

The phenomenon is best illustrated by first simulating multivariate functional data with all of the basis coefficients of the fixed effects fixed at $0$. Otherwise, we proceed as in the baseline simulation scenario, with $N = 280$, $n_i = 80$ and using the random effects parameters in Table \ref{tab:random-parameters}. We generate $500$ simulated datasets and, for each dataset, we estimate the first $10$ pooled mv-FPCs. Figure \ref{fig:eigenfunction-estimation-01} displays the $500$ estimates of each of the first three mv-FPCs in grey, with their empirical means overlaid as dashed black lines. The basis functions used to generate the data are indicated by the solid black lines. In this case, the basis coefficients are all marginally uncorrelated by construction and clearly the basis functions used to generate the data are, on average, being recovered as estimated mv-FPCs.

\begin{figure}
    \centering
    \includegraphics[page = 1, width = 1\textwidth]{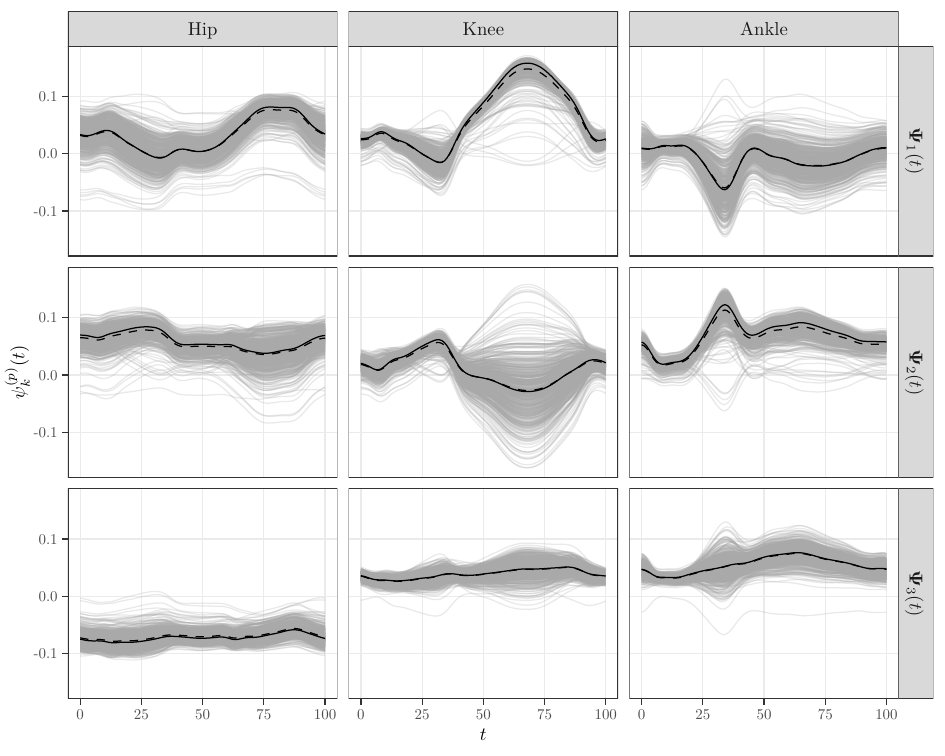}
    \caption{The results of the first simulation in Section \ref{app:eigenfunction-recovery}. The grey curves represent the mv-FPC estimates from $500$ simulated datasets generated by model \eqref{eq:data-generating-model} with the basis coefficients of the fixed effects fixed at $0$. The dashed black lines represent the averages of the 500 estimates. The solid black lines represent the true basis functions used to generate the data.}
    \label{fig:eigenfunction-estimation-01}
\end{figure}

We now repeat the experiment, but with the basis coefficients of the fixed effects set at their values in Table \ref{tab:fixed-parameters}, rather than being fixed at $0$. Again, we generate $500$ simulated datasets and on each one we estimate the first $10$ pooled mv-FPCs. Figure \ref{fig:eigenfunction-estimation-02} displays the results of this simulation. It can be seen that there are small discrepancies between the average mv-FPCs being recovered and the basis functions used to generate the data.
That is, the dashed and solid black lines do not match in certain parts of the functions (e.g., $\psi_2^{(knee)}(t)$).
This occurs because the data-generating models produce coefficients that are not marginally uncorrelated due to covariate effects. To demonstrate this, we calculate the marginal covariance matrix of the first ten basis coefficients via simulation. Then we compute its eigenvectors and use them to rotate the basis functions, to produce the eigenfunctions of the ``true" marginal covariance function. These functions are indicated by the solid red line in Figure \ref{fig:eigenfunction-estimation-02} and, as expected, are the average mv-FPCs being recovered in the simulation.

\begin{figure}
    \centering
    \includegraphics[page = 2, width = 1\textwidth]{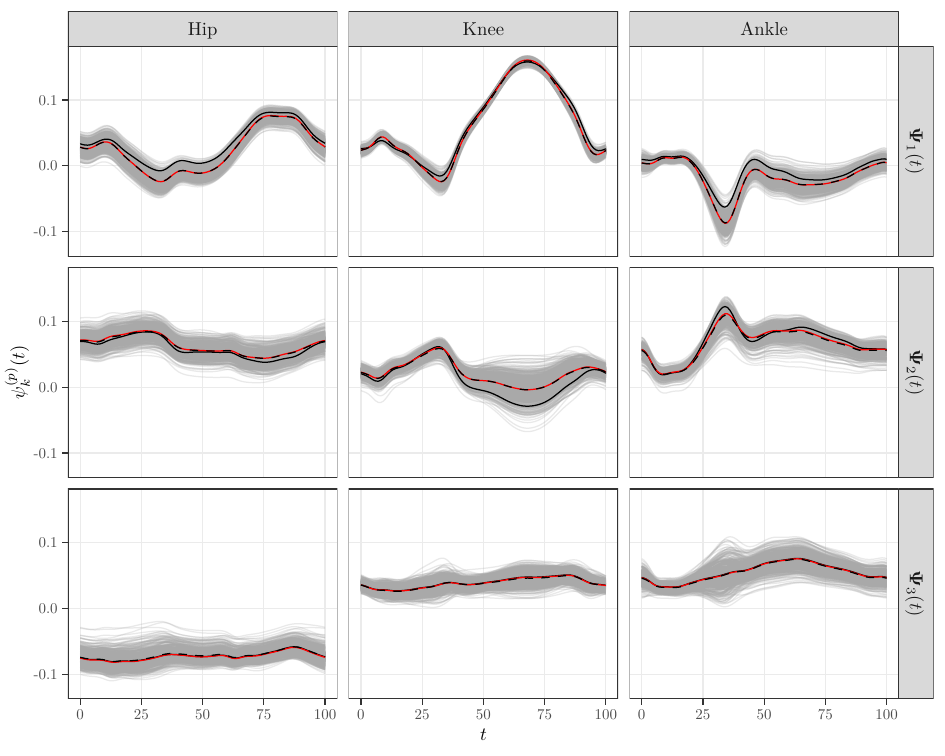}
    \caption{The results of the second simulation in Section \ref{app:eigenfunction-recovery}. The grey curves represent the mv-FPC estimates from $500$ simulated datasets generated by model \eqref{eq:data-generating-model} with the basis coefficients of the fixed effects set at their values in Table \ref{tab:fixed-parameters}. 
    The dashed black lines represent the averages of the 500 estimates. The solid black lines represent the true basis functions used to generate the data. The solid red lines represent linear combinations of the basis functions used to generate the data, given by an eigendecomposition of the marginal covariance matrix of basis coefficients.}
    \label{fig:eigenfunction-estimation-02}
\end{figure}

This is not a problem with estimation of the mv-FPCs or with data generation. The short simulation in this section has shown that we are, as expected, recovering the eigenfunctions of the marginal covariance function. The small marginal correlations among the basis coefficients are likely the result of simulating the covariates randomly from a distribution that mirrors the observed covariate distribution in our data application. 
The discrepancy is useful to note for designing future simulation studies in which eigenfunction estimation is used as an evaluation criteria.
A final point is that the first two eigenfunctions are estimated better in the second scenario than in the first, i.e., the grey curves in Figure \ref{fig:eigenfunction-estimation-02} exhibit less variability than those in Figure \ref{fig:eigenfunction-estimation-01}. This is simply an aretfact of setting the covariate effects to $0$ for the purpose of our demonstration -- the covariate effects account for a large amount of the variance in the mv-FPC1 coefficients, so setting them to $0$ reduces its overall variance explained. This reduces the difference between the eigenvalues associated with the first and second mv-FPCs. It is known that eigenfunctions become more difficult to estimate as their respective eigenvalues become less ``spread out" \parencite{reimherr_functional_2014}.

\subsection{Additional Simulation Results}\label{app:extra-simulation-results}

Figure \ref{fig:simulation-long-strength} displays the results of varying the strength of the longitudinal variation.
Computation times of the mv-FPCA step and the model fits (Figures \ref{fig:simulation-long-strength} (a) and (b), respectively) are relatively stable across the three levels.
Predictably, the difference in prediction error of individual observations between the naive model and the models that incorporate a longitudinal component (polynomial, spline and ml-FPCA) becomes more substantial as the strength of the longitudinal variation is increased (Figure \ref{fig:simulation-long-strength} (c)). This is because there is more longitudinal variation in the data that is not being captured by the naive model.
Overall, fixed effects estimation appears stable across all models and the three levels of longitudinal variation strength.
There does, however, appear to be more large outliers (i.e., simulation replicates with a large ISE) for the ml-FPCA model when the longitudinal variation is increased.
This may have to do with uncertainty in the estimated ml-FPCA basis functions being used.

\begin{figure}
    \centering
    \includegraphics[width = 1\textwidth]{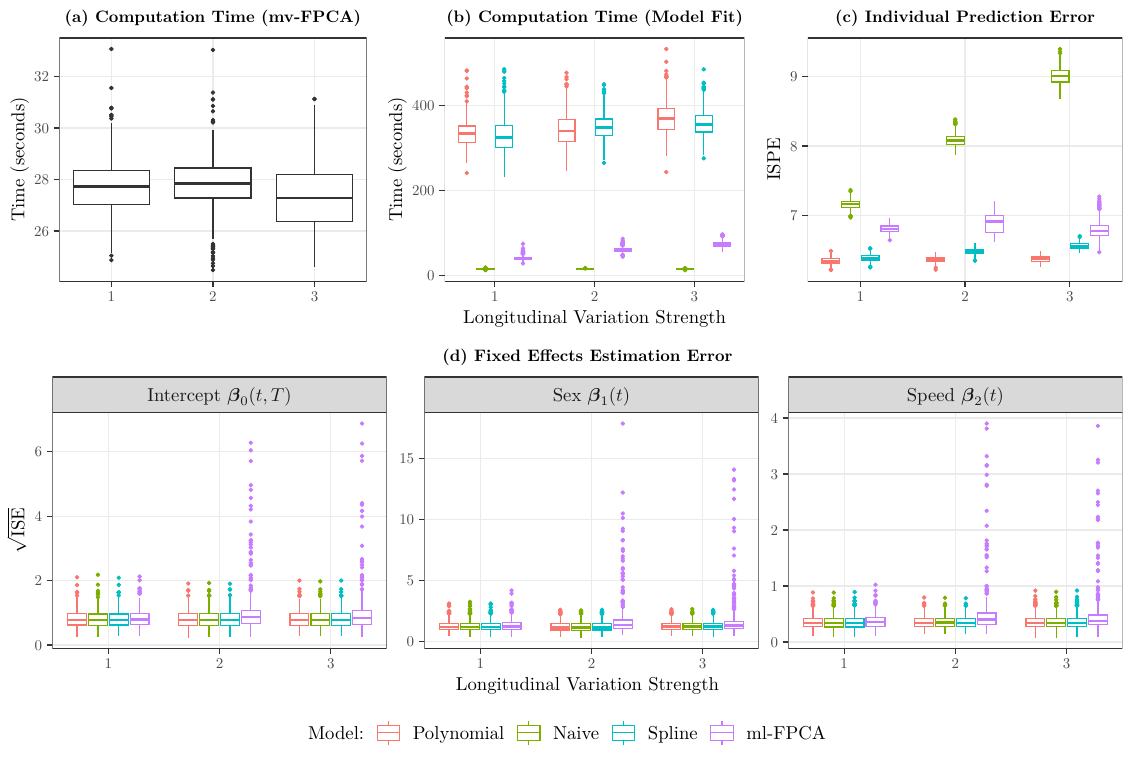}
    \caption{Results of the simulation varying the strength of the longitudinal variation. \textbf{(a)} The computation time for the mv-FPCA step in seconds. \textbf{(b)} The computation time for the model fits in seconds.
    \textbf{(c)} The integrated squared prediction error of held-out strides from the test set.
    \textbf{(d)} The integrated squared error of the fixed effects estimates.
    The number of subjects $N$ and the proportion of missing strides are fixed at their baseline values of $280$ and $0.1$, respectively.}
    \label{fig:simulation-long-strength}
\end{figure}

Figure \ref{fig:simulation-prop-missing} displays the results of varying the proportion of missing observations (i.e., strides) in the dataset.
The computation time for the mv-FPCA and model fits decreases as the number of missing strides is increased and the dataset used to fit the model becomes smaller (Figure \ref{fig:simulation-prop-missing} (a) and (b)).
Individual prediction errors increase as missingness is increased, reflecting that more observations per individual help in predicting the individual trajectories (Figure \ref{fig:simulation-prop-missing} (c)).
This effect is pronounced for the ml-FPCA model, possibly because more observations (per individual) are needed to obtain better estimates of the ml-FPCA basis functions. The estimates of the fixed effects are stable across the levels of missingness (Figure \ref{fig:simulation-prop-missing} (d)).

\begin{figure}
    \centering
    \includegraphics[width = 1\textwidth]{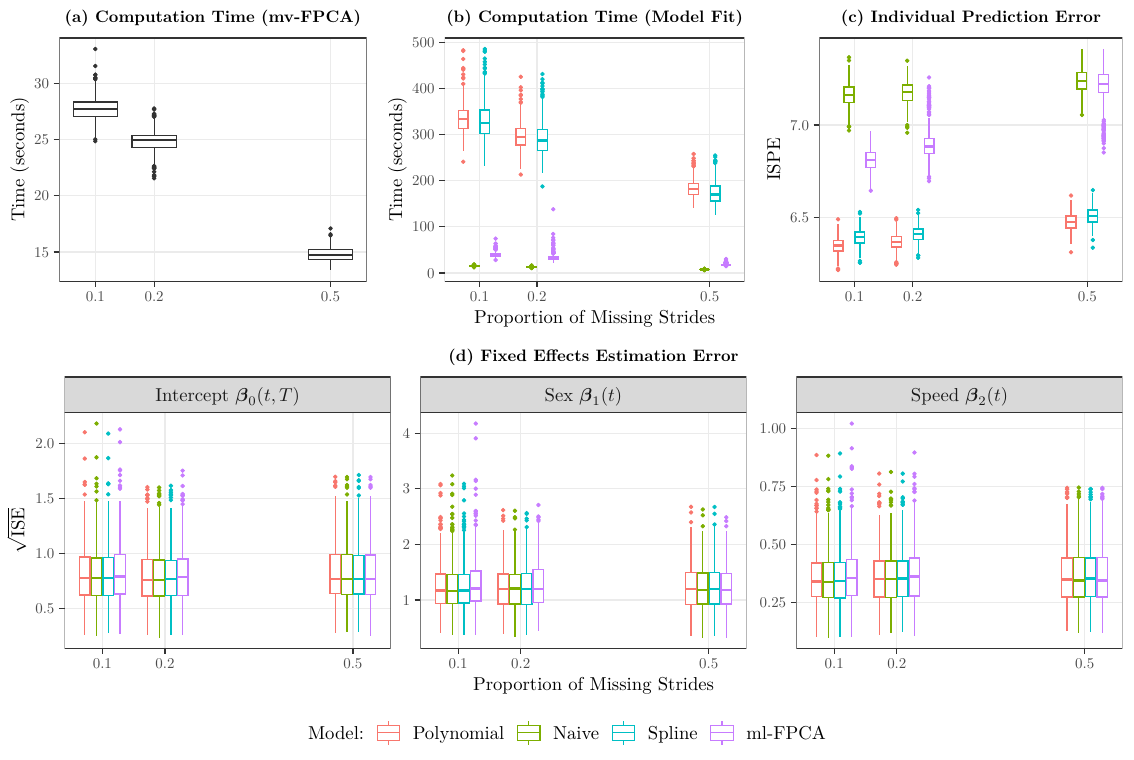}
    \caption{Results of the simulation varying the proportion of missing data used in model fitting. \textbf{(a)} The computation time for the mv-FPCA step in seconds. \textbf{(b)} The computation time for the model fits in seconds.
    \textbf{(c)} The integrated squared prediction error of held-out strides from the test set.
    \textbf{(d)} The integrated squared error of the fixed effects estimates.
    The number of subjects $N$ and the strength of the longitudinal variation are fixed at their baseline values of $280$ and $1$, respectively.}
    \label{fig:simulation-prop-missing}
\end{figure}

Table \ref{tab:singularity-table} displays results of checking singular fit warnings in the scalar mixed models fitted to the mv-FPCA scores. The averages are taken over the first $10$ scores and over the $500$ simulation replicates in each scenario (when more than $10$ mv-FPCs were estimated, those beyond the $10$th were not included in this check). The naive model is not included here because none of the fits were singular.
Singular fit issues only appear to be a problem for the spline model, possibly because the largest unstructured $\mathbf{S}^*$ matrix is being estimated and the spline basis functions are not orthogonal. For this model, the proportion of fits that were singular ranges between $0.3$ and $0.45$ and appears to depend on the simulation scenario.

% latex table generated in R 4.1.2 by xtable 1.8-4 package
% Mon Aug 21 09:31:36 2023
\begin{table}[ht]
\centering
\begin{tabular}{llllll}
  \toprule
{\bfseries \small $N$} & {\bfseries \small Pr. Missing} & {\bfseries \small Lon. Strength} & {\bfseries \small Polynomial} & {\bfseries \small Spline} & {\bfseries \small ml-FPCA} \\ 
  \midrule
280 & 0.1 & 1 & 0.01 (0.001) & 0.357 (0.007) & 0 \\ 
  500 & 0.1 & 1 & 0.004 (0.001) & 0.338 (0.007) & 0 \\ 
  1000 & 0.1 & 1 & 0.001 ($<$ 0.001) & 0.308 (0.007) & 0 \\ 
  280 & 0.2 & 1 & 0.012 (0.002) & 0.318 (0.007) & 0.001 ($<$ 0.001) \\ 
  280 & 0.5 & 1 & 0.036 (0.003) & 0.303 (0.006) & 0 \\ 
  280 & 0.1 & 2 & 0.002 (0.001) & 0.403 (0.007) & 0 \\ 
  280 & 0.1 & 3 & 0.02 (0.002) & 0.447 (0.007) & 0 \\ 
   \bottomrule
\end{tabular}
\caption{Proportion of singular fit warnings from the model fits. In cases where the proportion is non-zero, a Monte Carlo standard error estimate for the true proportion is reported in brackets to convey uncertainty due to the finite number of simulations.} 
\label{tab:singularity-table}
\end{table}

\clearpage
\section{Additional Results}\label{app:additional-results}

\subsection{Data Preparation}

Figure \ref{fig:data-prep-plot} (a) displays a histogram of the treadmill run duration (i.e., capture period) variable that was used to perform the subject-specific normalisation of the longitudinal time variable $T$.
The average duration was exactly 60 seconds (dashed black line) and $93\%$ of the durations were between $50$ and $70$ seconds (dotted lines).
Figure \ref{fig:data-prep-plot} (b) displays a histogram of the number of strides on each side for every subject included in the analysis. This quantity varies because subjects take differing numbers of strides during the treadmill run and also because subjects had strides removed due to data-collection errors.

\begin{figure}[h]
    \centering
    \includegraphics[width = 0.9\textwidth]{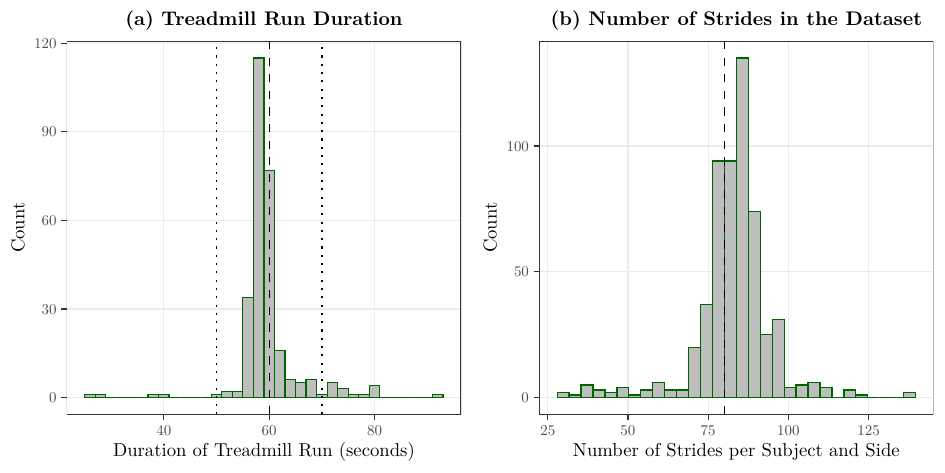}
    \caption{\textbf{(a)} Histogram of the treadmill run duration of subjects included in the analysis. The dashed vertical line indicates $60$ seconds and the vertical dotted lines indicate $50$ and $70$ seconds. \textbf{(b)} Number of strides per subject and side for subjects included in the analysis. The dashed vertical line is at $80$.}
    \label{fig:data-prep-plot}
\end{figure}

Figure \ref{fig:basis-transformation} (a) and (b) display a scree-plot and the cumulative percentage of variance explained, respectively, for the mv-FPCA.
To graphically assess the mv-FPCA reconstruction \parencite{morris_comparison_2017}, Figure \ref{fig:basis-transformation} (c) displays five randomly-selected observations from the test set. The mv-FPC reconstruction appears to reconstruct the functions well. 

\begin{figure}[h]
    \centering
    \includegraphics[page = 3, width = 1\textwidth]{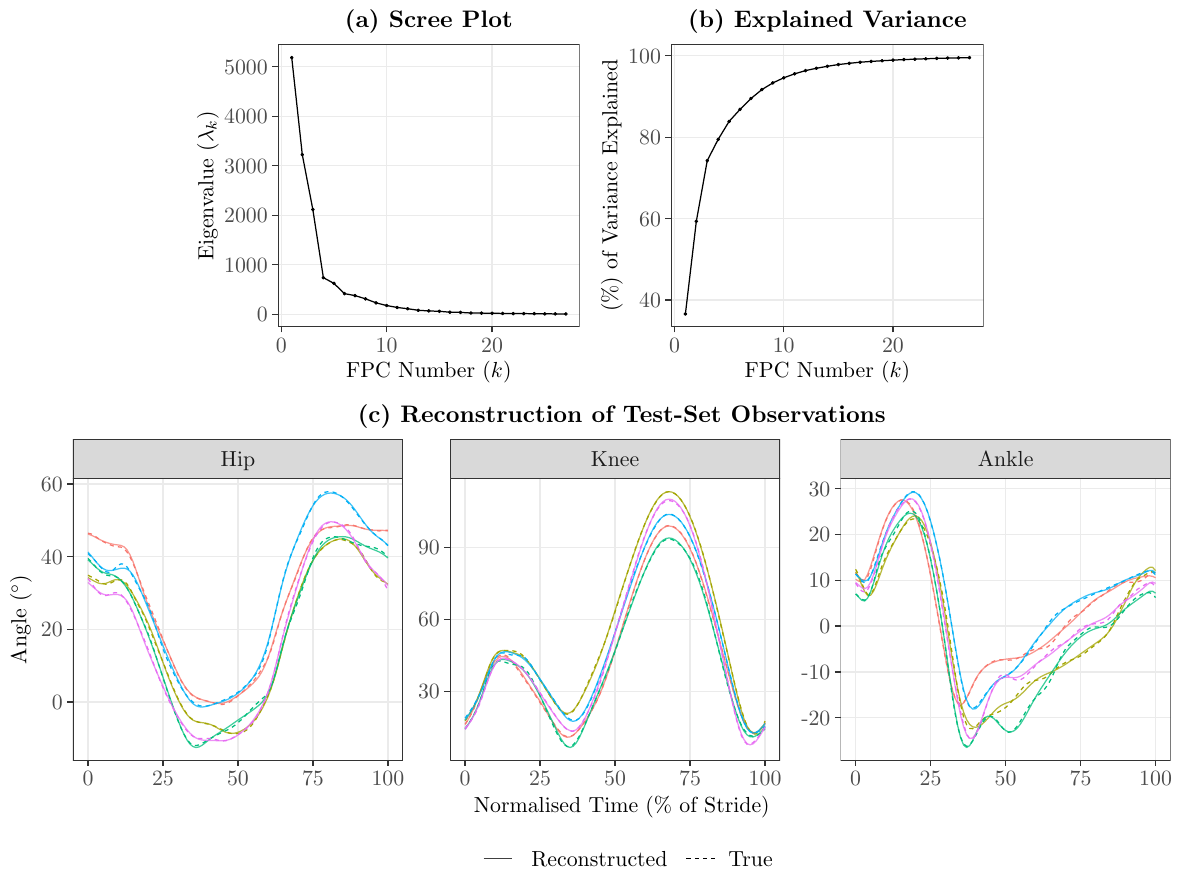}
    \caption{Results of the mv-FPCA representation of the multivariate functional data. \textbf{(a)} Scree plot displaying the eigenvalue $\lambda_k$ associated with each retained mv-FPC $k = 1, \dots, 27$. \textbf{(b)} The cumulative percentage of variance explained by each each mv-FPC. \textbf{(c)} A random sample of five multivariate functional observations from the test set (solid lines) and their mv-FPCA reconstructions (dashed line).}
    \label{fig:basis-transformation}
\end{figure}

\subsection{Fixed Effects} \label{app:fixed-effects}
As the intercept $\boldbeta_0(t, T)$ is modelled as a linear combination of four natural cubic B-spline basis functions in the longitudinal direction, we simply inspect the four regression coefficient functions associated with these longitudinal basis functions separately. An approach of this type was first employed by \textcite{park_longitudinal_2015}, as it alleviates the need to inspect uncertainty estimates of the full two-dimensional surface.
Figure \ref{fig:spline-coefficients} displays the estimated regression coefficient functions of the natural cubic B-spline basis functions used to model the intercept in the longitudinal direction. 
The black solid line represents the point estimate, the dotted black lines indicate pointwise $95\%$ confidence intervals and the light blue ribbons represent $95\%$ simultaneous confidence bands.

The estimated coefficients are small in magnitude (almost all $< 1^\circ$). In the hip and knee dimensions, the simultaneous confidence bands contain $0$ for almost all $t$. For the ankle dimension, the bands do not contain $0$ for a short period around $t = 75\%$. However, the range of values for the effect that are captured by the band are still very small. 
To emphasise this, we calculate the estimated intercept function $\widehat{\boldsymbol{\beta}}(t, T)$ on an equidistant grid of longitudinal time points along $[0,1]$ and display the results on a rainbow plot in Figure \ref{fig:rainbow-intercept}.
Figure \ref{fig:rainbow-intercept} (a) displays the estimated intercept function on its original scale, where the longitudinal effects are not visible. 
Figure \ref{fig:rainbow-intercept} (b) displays the estimated intercept centred around the overall mean function, allowing the longitudinal effect to be seen more clearly. In particular, the effect in the ankle at about $t = 75\%$ corresponds to a change of $< 0.5^\circ$ over the course of the treadmill run. 
Overall, we can conclude that changes in the longitudinal direction, although statistically significant in certain places, are minimal and the intercept function is approximately constant along $T$.

\begin{figure}
    \centering
    \includegraphics[width = 0.95\textwidth]{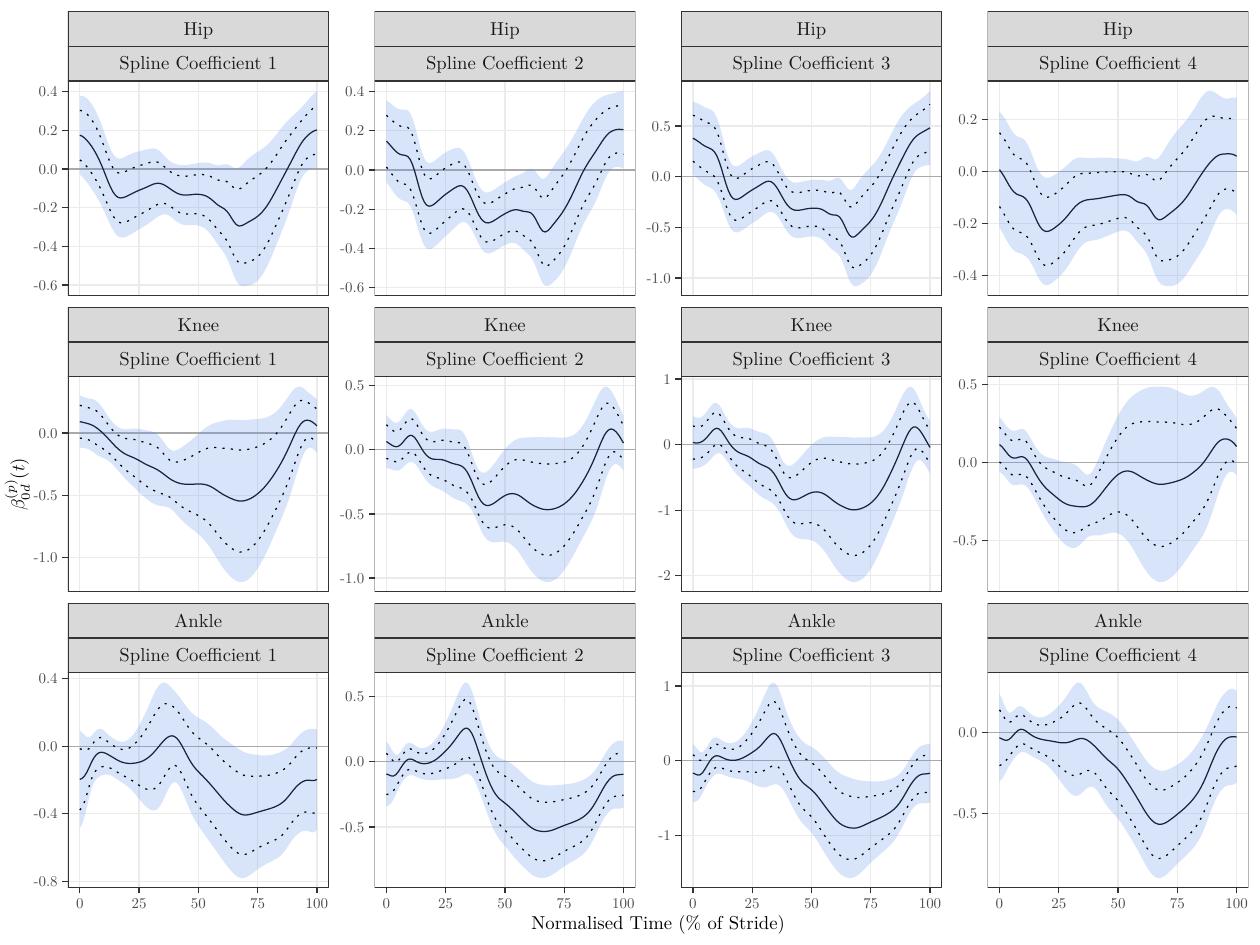}
    \caption{The regression coefficient functions of the natural cubic B-spline basis functions used to model the intercept in the longitudinal direction. The black solid line represents the point estimate, the dotted black lines indicate pointwise $95\%$ confidence intervals and the light blue ribbons represent $95\%$ simultaneous confidence bands.}
    \label{fig:spline-coefficients}
\end{figure}

\begin{figure}
    \centering
    \includegraphics[width = 0.95\textwidth, page = 7]{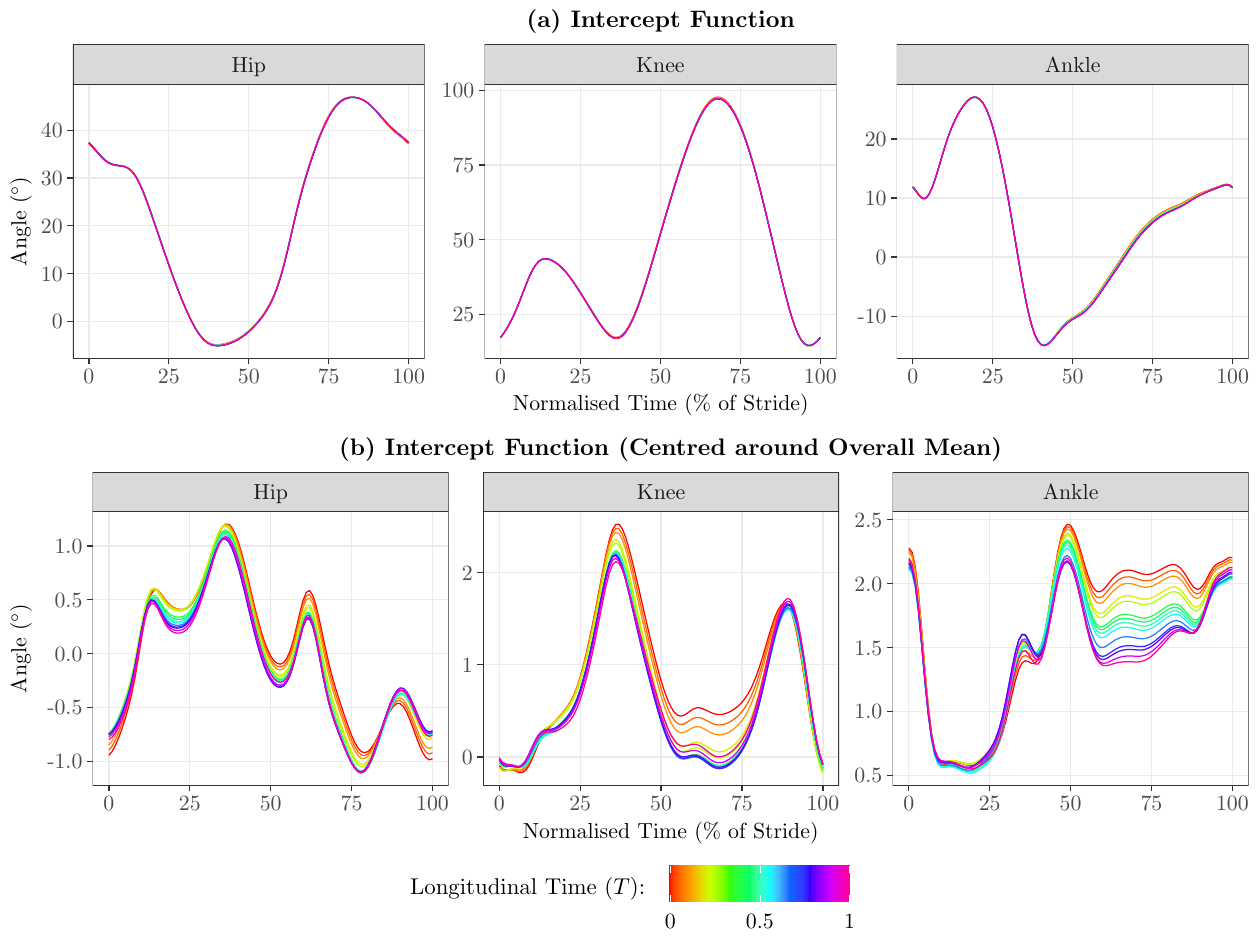}
    \caption{Rainbow plot of the estimated intercept function where colouring indicates the longitudinal time $T$. \textbf{(a)} The estimated intercept function. \textbf{(b)} The estimated intercept function centred around the overall mean function.}
    \label{fig:rainbow-intercept}
\end{figure}

\subsection{Diagnostics}\label{sec:lfmm-diagnostics}

Figure \ref{fig:diagnostics-mvfpc-1} displays residual diagnostics from the (spline) scalar linear mixed effects model fitted to the first mv-FPC. The BLUPs of the subject and subject-and-side level random intercepts appear to be approximately Gaussian distributed (panels (a) and (b)). 
The conditional residuals appear to be symmetrically distributed with heavier tails than a Gaussian distribution as evidenced by the departures from the straight line at each end of the Gaussian quantile-quantile (Q-Q) plot (panel (c)). 
The residual autocorrelation function (ACF) demonstrates that residual autocorrelation is significantly reduced by the longitudinal spline model relative to the naive model (panel (d)). 
Though there is still evidence of some autocorrelation at shorter lags indicating that adding an AR(1) residual correlation structure may still be beneficial, the longitudinally varying terms at the subject level appear to have captured the majority of the correlation.

\begin{figure}
    \centering
    \includegraphics[width = 0.75\textwidth]{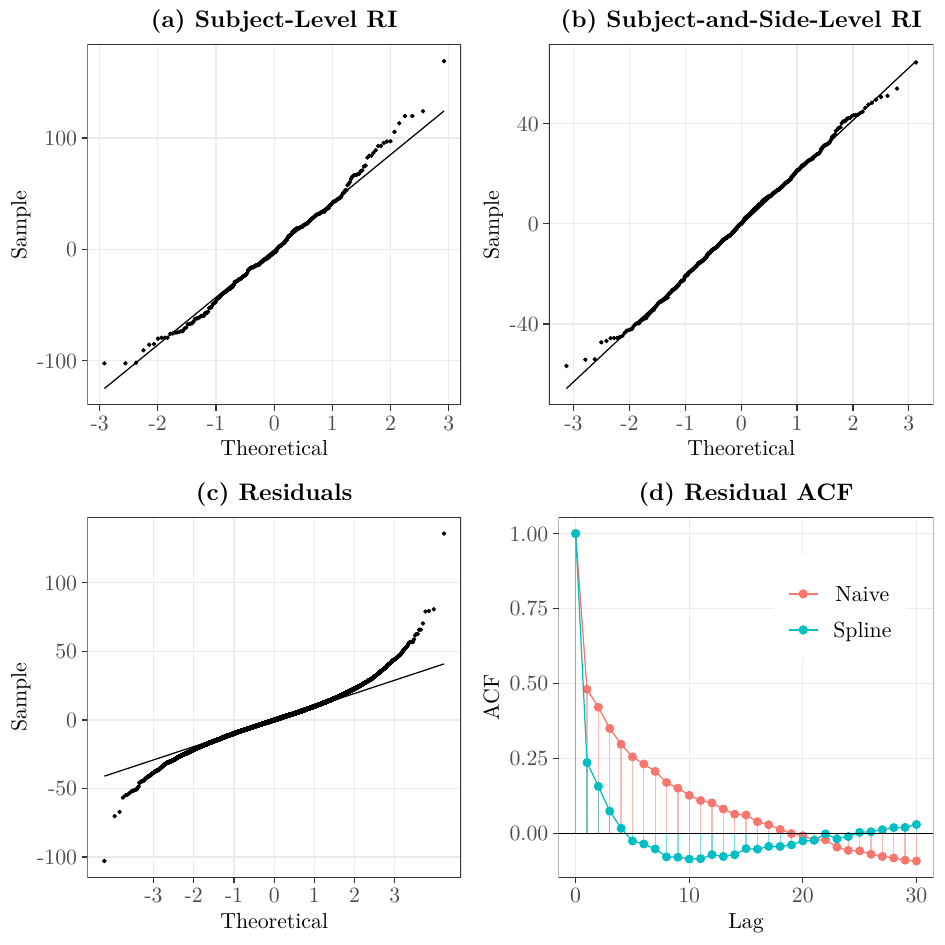}
    \caption{Residual diagnostics for the (spline) scalar linear mixed model fitted to mv-FPC1. \textbf{(a)} A Gaussian quantile-quantile (Q-Q) plot of the BLUPs of the subject-level random intercept. \textbf{(b)} A Gaussian Q-Q plot of the BLUPs of the subject-and-side-level random intercept. \textbf{(c)} A Gaussian Q-Q plot of the conditional residuals. \textbf{(d)} The residual autocorrelation (ACF) function of the conditional within-subject-and-side residuals from both the spline and naive models. }
    \label{fig:diagnostics-mvfpc-1}
\end{figure}

\end{document}